\documentclass[useAMS,usenatbib,usegraphicx]{mn2e}
\usepackage{amsmath,amssymb}
\usepackage{graphicx}  
\usepackage{hyperref}

\usepackage[usenames,dvipsnames]{xcolor}
\newcommand\beq{\begin{equation}}
\newcommand\eeq{\end{equation}}
\usepackage[margin=10pt,font=small,labelfont=bf]{caption}
\usepackage{units}
\usepackage{eucal}
\usepackage{textcomp}
\usepackage{placeins}
\usepackage{grffile}

\def\gsim{ \lower .75ex \hbox{$\sim$} \llap{\raise .27ex \hbox{$>$}} }
\def\lsim{ \lower .75ex\hbox{$\sim$} \llap{\raise .27ex \hbox{$<$}} }

\newcommand{\mdisc}      {\ensuremath{M_{\mathrm{Disc}}}}

\newcommand{\mbhb}     {\textcolor{Plum}{{\ensuremath{~M_{\mathrm{BHB}}}}}}
\newcommand{\mbhbdot}     {\textcolor{Plum}{{\ensuremath{~\dot{M}_{\mathrm{BHB}}}}}}
\newcommand{\ratio}     {\ensuremath{ {  \textcolor{RoyalPurple}{\dot{M}_2}}/{ \textcolor{orange}{\dot{M}_1}  } }}

\newcommand{\meins}      {\ensuremath{~M_{\mathrm{BH}_1}}}
\newcommand{\mzwei}      {\ensuremath{~M_{\mathrm{BH}_2}}}
\newcommand{\mein}      {\ensuremath{M_{1}}}
\newcommand{\mzwe}      {\ensuremath{M_{2}}}
\newcommand{\ecc}      {\textcolor{OliveGreen}{\ensuremath{\mathfrak{e}}}}

\renewcommand{\a}      {\textcolor{blue}{\ensuremath{\mathfrak{a}}}}
\newcommand{\mmu}      {\textcolor{Orange}{\mu}}
\newcommand{\q}      {\ensuremath{\mathfrak{q}}}
\renewcommand{\i}      {\textcolor{Maroon}{\ensuremath{\mathrm{i}}}}

\newcommand{\G}       {\ensuremath{\mathrm{G}}}

\newcommand{\tordens}   {d$T_{{\rm G},z}/$d$r $\,}
\newcommand{\gas}       {\ensuremath{\mathcal{G}}}

\newcommand\bea{\begin{eqnarray}}
\newcommand\eea{\end{eqnarray}}

\title[Retrograde discs around SMBHBs]
{Migration of  massive black hole binaries in self--gravitating accretion discs: Retrograde versus prograde}

\author[C. Roedig and A. Sesana]{Constanze Roedig$^{1,2}$\thanks{E-mail: croedig@aei.mpg.de} and Alberto Sesana$^1$ \\
$^{1}$ Max-Planck-Institut f{\"u}r Gravitationsphysik, Albert Einstein
Institut, Am M{\"u}hlenberg 1, 14476 Golm, Germany \\
$^{2}$ Department for Physics and Astronomy, The Johns Hopkins University, 3400 N Charles Street, Baltimore, MD, 21218, USA
}

\begin{document}
\date{Received ---}

\pagerange{\pageref{firstpage}--\pageref{lastpage}}
\pubyear{2013}
\maketitle

\begin{abstract}
We study the interplay between mass transfer, accretion and gravitational torques onto a black hole binary migrating
in a self-gravitating, retrograde circumbinary disc. A direct comparison with an identical prograde disc shows that: (i)
because of the absence of resonances, the cavity size is a factor \a(1+\ecc) smaller for retrograde discs; (ii)
nonetheless the shrinkage of a circular binary semi--major axis \a\, is identical in both cases; (iii)
a circular binary in a retrograde disc remains circular while eccentric binaries grow more eccentric.
For non-circular binaries, we measure the orbital decay rates and the eccentricity growth rates to be exponential as long as the binary orbits in the plane of its disc. Additionally, for these co-planar systems, we find that interaction ($\sim$ non--zero torque) stems only from the cavity edge plus \a(1+\ecc)  in the disc, i.e. for dynamical purposes, the disc can be treated as a annulus of small radial extent. We find that simple 'dust' models in which the binary-
disc interaction is purely gravitational can account for all main numerical  results,
both for prograde and retrograde discs.
Furthermore, we discuss the possibility of an instability occurring for highly eccentric binaries causing
it to leave the disc plane, secularly tilt and converge to a prograde system.
Our results suggest that there are two stable configurations for binaries in self-gravitating discs:
the special circular retrograde case and an eccentric (\ecc$\sim$ 0.6) prograde configuration as a stable attractor.
 
\end{abstract}

\section{Introduction}
\label{intro}
Supermassive black hole binaries (BHBs){\footnote{Throughout the paper we always refer to massive/supermassive binaries, we therefore drop the adjective and simply refer to BHBs.}} have been predicted to be ubiquitous intermediate products of comparable mass galaxy mergers \citep[e.g.][]{begelman80}. We consider the possibility of a BHB that has migrated to reach parsec separation and is now evolving in a retrograde selfgravitating disc. Without making any a-priori statements about the astrophysical likelihood of this scenario, we compare the migration in a retrograde disc  to the prograde case. Following the seminal work of \cite{GoldreichTremaine80}
in the context of planetary dynamics, several studies of sub-parsec BHBs in gaseous discs \citep{Artymowicz1994,syer95,arty96,Ivanov99, Gould2000,Armitage:2002,Armitage:2005,hayasaki07,haya08,MacFadyen2008,hayasaki09,Jorge09,Haiman2009, Lodato2009,LinPapa11, nixon11a,Roedig2011,Shi11} employing various techniques, have  established  the picture sketched below. 
Initially the binary migrates inwards following the 'fluid elements' of the discs (type I) migration. 
When the local mass of the disc \citep[see, e.g.,][for a thoughtful discussion]{Kocsis2012} is of the order of the mass of the lighter hole, torques exerted by the binary are strong enough to evacuate the central region of the disc carving a central cavity. Accretion continues onto both holes through the cavity. In the prograde case (studied by the aforementioned papers), the cavity itself has a radius which is about twice the binary separation, and is held up by outer Lindblad resonances. The gas at the inner edge of the circumbinary disc leaks through the cavity streaming towards either holes. The streams bend in front of each hole and are partially gravitationally captured forming {\it minidiscs}. However, a large amount of gas is flung back out impacting the disc edge, efficiently mediating the disc-binary energy and angular momentum exchange \citep{MacFadyen2008,Shi11}. The semi--major axis, \a, shrinks 
and the eccentricity, \ecc, grows and saturates around \ecc$\sim$ 0.6 \citep{Roedig2011}.

Recently \cite{nixon11a} investigated the dynamical interaction of BHB embedded in light, non-selfgravitating circumbinary retrograde discs.
The absence of outer Lindblad resonances (OLR) allows the circumbinary disc to survive unperturbed outside the orbit of the binary, 
with the inner edge impacting directly onto the secondary BH. So, the retrograde cavity is smaller and there is neither room to accommodate
large {\it minidiscs} nor will there be {\it stream bending} nor will there be sling--shots. The gas at the inner edge is either accreted
directly by the secondary hole, or loses its energy and angular momentum in the impact being funneled onto the primary. 
We study here the evolution of BHB embedded in massive, self gravitating retrograde discs.
The first goal of this paper is to compare the evolution of the orbital elements of an initially circular BHBs migrating inside pro-- versus 
retrograde discs. The pro-retro run comparison allows us to better understand the physics of the BHB-disc interaction. We 
demonstrate that, in both configurations, the evolution of the binary elements is well reproduced if the BHB-disc interplay is described
in terms of purely gravitational interactions, in which the gas particles are effectively treated as dust. In this picture, the hydrodynamical properties of the disc is only relevant in determining the pace at which gas is supplied for direct interaction with the BHB. 
At very high eccentricity we find hints of an instability that leads to the binary tilting out of the rotation plane of the disc.
We pursue this possibility with a dedicated higher resolution simulation and point out the most likely cause of this behavior.
Combining these finding with those of \cite{Roedig2011}, we discuss that BHBs in self-gravitating retrograde discs have three possible
end-stages: (1) a stable circular motion coupled with a orbit decay similar to the aligned state; 
 depending on the stochastic occurrence of the instability either (2)
a fast inspiral if the exponential eccentricity growth and semi--major axis decay bring the BHB into the gravitational wave regime
or (3) re--alignment of the BHB with its disc and evolution towards the eccentric, prograde standard scenario.
 
The paper is organized as follows. In \S ~\ref{methods}, we describe the numerical method, initial conditions 
and parameters use in our simulations. In \S~\ref{sec:analysis}, we provide a compilation of our numerical findings,
that we try to explain coherently with a simple model presented in \S \ref{sec:analytics}.  
The aforementioned tilting instability is studied in \S~\ref{sec:tilting}. 
Concluding, we discuss and summarize the overall findings in \S~\ref{discussion}.

\section{Method}
\label{methods}

\begin{table}
  \caption{
{\bf \sc Run labels and initial parameters in code units}. The prograde run is taken from paperI \citet{Roedig2012}. On a side note: $N$ is the resolution, ${\tt ^{**}}$ denotes that additional cross--checks were preformed to ensure that all conservation laws were numerically satisfied.
\label{initretro}}
\begin{center}
\begin{tabular}{lcccccc}
\hline 
\hline
Model \quad & $\a_0$ &  $\ecc_0$   &$r_{\rm sink} $  \quad& N & $\mdisc/\mbhb$ \\
\hline
$\mathtt{pro\, \ecc=0}$  \quad&$1.0$   & $0.02$ \quad& $0.05 $& 2 M&   $0.186 $\\
\hline
$\mathtt{count00}$     \quad&$0.97$ & $0.02$ \quad & $0.05$ & 3 M&   $0.13 $\\
$\mathtt{count01}$     \quad&$0.92$ & $0.09$ \quad & $0.05$ & 3 M&   $0.13 $\\
$\mathtt{count02}$     \quad&$0.83$ & $0.19$ \quad & $0.05$ & 3 M&   $0.13 $\\
$\mathtt{count03}$     \quad&$0.74$ & $0.30$ \quad & $0.05$ & 3 M&    $0.13 $\\
\hline
$\mathtt{retro\, \ecc=0}$  \quad&$1.0$   & $0.02$ \quad& $0.05 $& 2 M&    $0.186 $\\
$\mathtt{retro\, \ecc=0.1}$  \quad&$1.0$   & $0.10$ \quad& $0.05 $& 2 M&    $0.186 $\\
$\mathtt{retro\, \ecc=0.3}$  \quad&$1.0$   & $0.27$\quad & $0.05 $& 2 M&    $0.186$\\
$\mathtt{retro\, \ecc=0.5}$  \quad&$1.0$   & $0.46$\quad & $0.05 $& 2 M&     $0.186$\\
$\mathtt{retro\, \ecc=0.8}$  \quad&$1.0$   & $0.83$\quad & $0.05 $& 2 M&    $0.186$\\
\hline
$\mathtt{retro^{**} \ecc=0.7}$  \quad&$1.0$   & $0.56$\quad &$ 0.05 $& 4 M&$0.176$\\ 
$\mathtt{retro^{**} \ecc=0.9}$  \quad&$1.0$   & $0.94$\quad & $0.04$ & 4 M&$0.186$\\   
\hline
\hline 
\end{tabular}
\end{center}
\end{table}
The model and numerical setup of this work is closely related to that of \cite{Roedig2011,Roedig2012}(the latter hereafter paperI) and \cite{Jorge09}, hence we only outline the key aspects in the following, and refer the reader to these  papers for further details.

We simulate a self-gravitating gaseous disc of mass $\mdisc$ around two BHs of combined mass $\mbhb=M_1+M_2$, mass ratio $\q=M_2/M_1 \sim 1/3$,
eccentricity $\ecc$ and semi-major axis $\a$, using the SPH-code {\sc Gadget-2} \citep{Springel05}
in a modified version that includes sink-particles which model accreting BHs \citep{Springel05b, Cuadra2006a}. 
The disc, which is counter-rotating with the BHs, radially extends to about $7\a$, contains a circumbinary cavity
of radial size $\gsim\a$ and is numerically resolved by at least $2$ million particles. 
The numerical size of each BH is denoted as $r_{\rm sink}$, the radius below which a particle is accreted,
removed from the simulation, and its momentum is added to the BH \citep{Bate95}.
The gas in the disc is allowed to cool on a time scale proportional to the local dynamical time of the disc $t_{\rm dyn}=2 \pi / \Omega_0$,
where $\Omega_0 =(\G M_0/\a_0^3)^{1/2}$ is the initial orbital frequency of the binary\footnote{Subscripts $0$ refer to the initial values 
of any parameter.}. To prevent it from fragmenting, we force the gas to cool slowly, setting $\beta = t_{\rm cool}/t_{\rm dyn} = 10$ \citep{Gammie01,Rice05}.   This choice of $\beta$ is discussed in previous work (e.g. paperI). All figures are shown in code units $\G= M_0=1$.

\subsection{Discussion of initial conditions}
\label{subsec:ICs}
 
 We start the simulations using two different approaches for obtaining initial conditions (ICs), the relevant quantities of
 which are listed in Table~\ref{initretro}. In the standard case, labeled ${\tt retro}$ throughout,
 a relaxed prograde disc is taken and the orbital angular momentum vector of the BHB, ${\bf L}_{\rm BHB}$,
 is flipped and the eccentricity is changed. To be specific, we took the ICs of model {\it adia05} in paperI, 
 in order to facilitate a direct prograde vs retrograde comparison. This prograde run is labeled ${\tt pro\,\ecc=0}$. 
 Additionally, to measure how fast the memory of any out-of-equilibrium IC is lost and in order to avoid any spurious systematics,
 the following test was performed: the cavity was filled with gas to form a solid accretion disc lacking any low density region around the BHB. A circular BHB was placed around the center of mass (CoM) and integrated to measure how many orbits the system would need to relax. We observed an initial transient marked by high accretion and BHB shrinkage 
that lasted $45$ orbits. Thereafter the system converged to a perfectly relaxed behavior,
the evolution of which in terms of {\it time derivative} of the {semi--major axis $\dot{\a}$},
the {eccentricity $\dot{\ecc}$} and mass, $\mbhbdot$, could not be distinguished from the ${\tt retro\, \ecc= 0}$ setup.
We thus only show data that exclude the first $45$ orbits, to avoid any potential initial transient. Unless otherwise stated, 
for each setup we use the orbits $45-90$ for the analysis.

The second set of ICs, labeled ${\tt count}$, uses a disc that was left to relax onto a retrograde, circular BHB for $750$ orbits,
which is why the disc is lighter as it got depleted over time. After 750 orbits, we modify the binary semi--major axis and eccentricity,
keeping the quantity \a(1+\ecc) approximately constant.
This is physically motivated by the previous observation that the gap size in a prograde disc is, at zero order,
determined by the apocenter of the BHB orbit\footnote{In reality, in the prograde case,
we observed the gap size to be slightly larger than 2\a(1+\ecc) due to non-accreted gas being flung out
in a sling--shot, impacting at the gap edge, shocking and pushing the edge further outwards.
In the retrograde setup, we do not expect sling-shots to play any role.}.
We then let the system evolve for 90 orbits and, as in the ${\tt retro}$ case, we keep orbits $45-90$ for the analysis.

\begin{figure}
\includegraphics[width=0.9\linewidth]{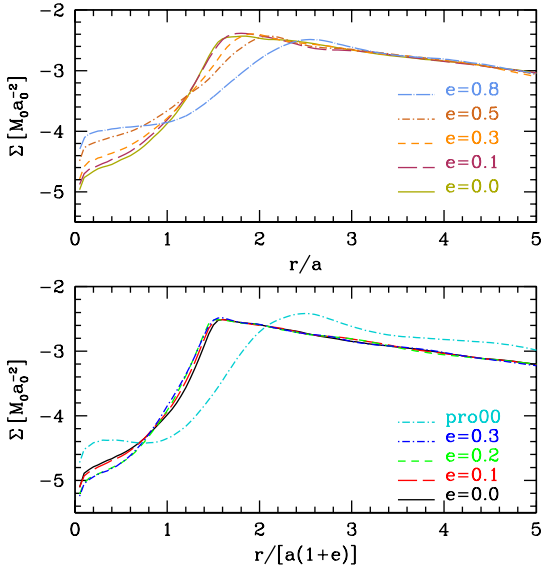}
\caption{
{\sc Retrograde runs, average surface density $\Sigma(r)$} - {\it Top:} $"{\tt retro}$ runs. {\it Bottom:} constant $(1+\ecc)\a$ runs  $"{\tt count}''$ runs, Additionally, the circular prograde case is overlayed, in which the surface density peaks at a radius that is 1 \a\, larger than in the retrograde case. Note that \mdisc\ in the ${\tt count}$ runs is smaller than in the ${\tt retro}$ runs. See text for more details; quantities are in code units.
}
\label{countsigma}
\end{figure}
\begin{figure}
\includegraphics[width=0.9\linewidth]{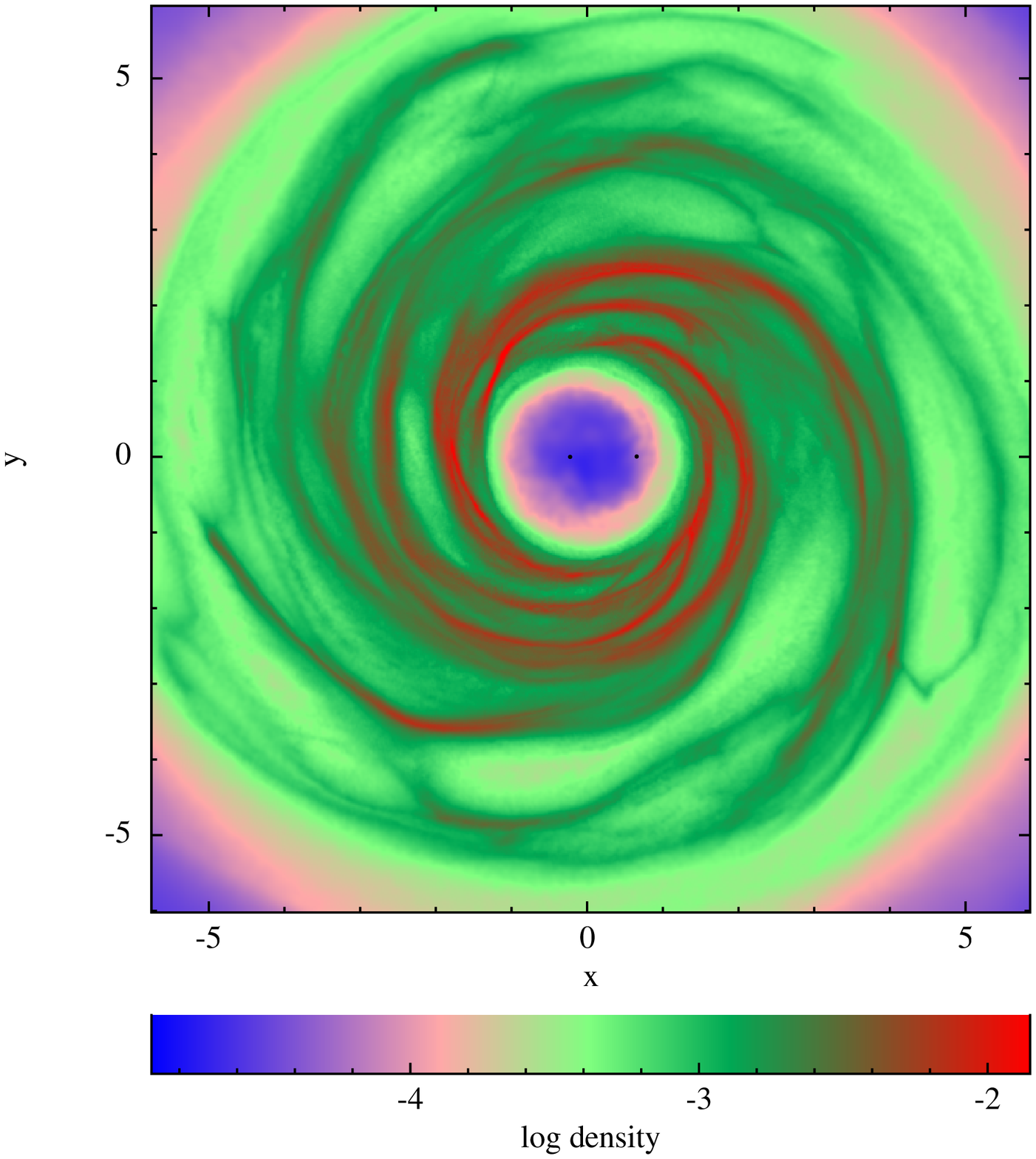}
\includegraphics[width=0.9\linewidth]{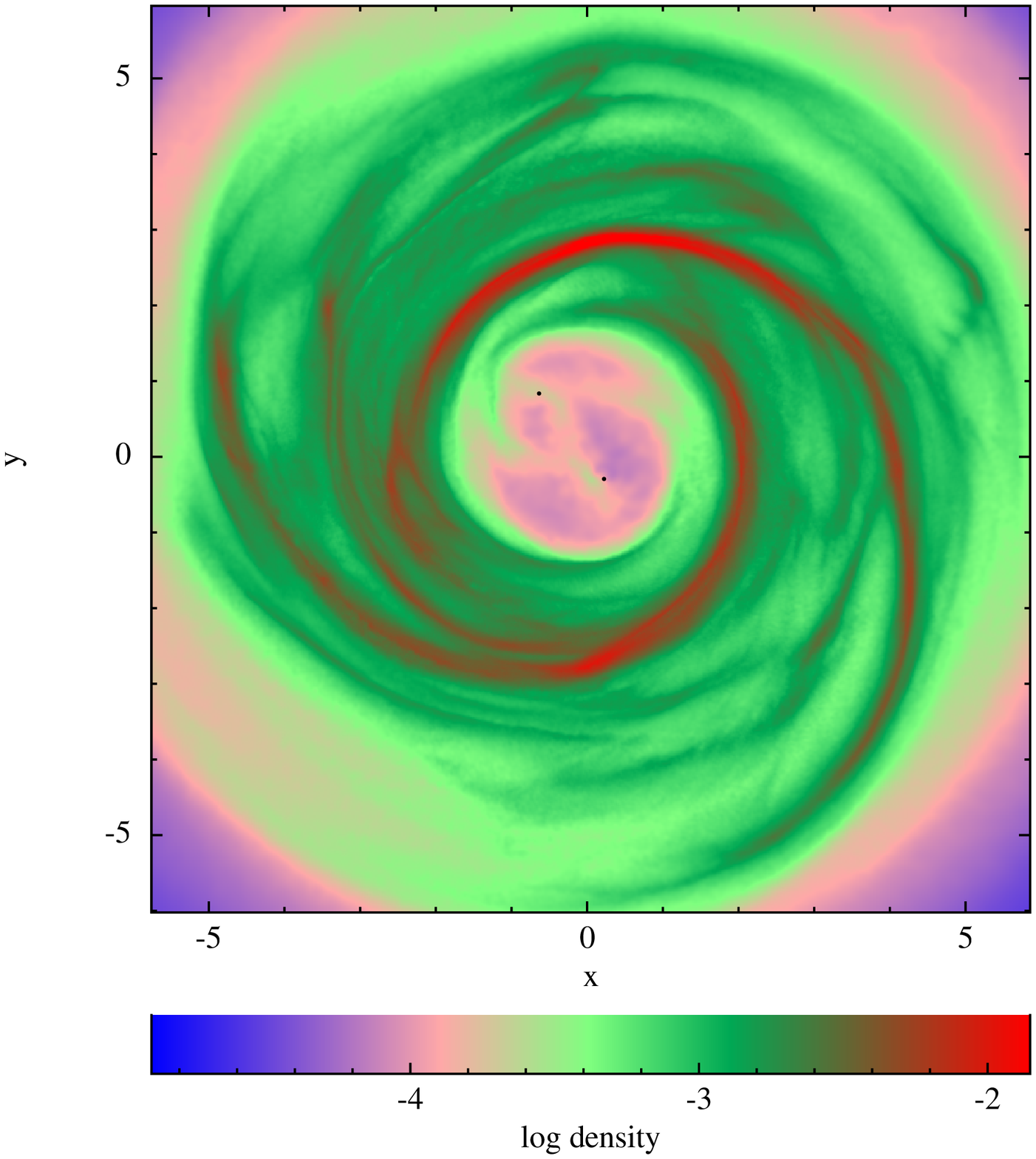}
\caption{
{\sc Disc surface density of runs ${\tt retro\,\ecc= 00}$ (top) and ${\tt retro\,\ecc= 08}$ (bottom)} - Colors are in log-density scale as indicated by the colour-bars, all plots are in code units $\G=M_0=\a_0=1$ and snapshots are taken at the same time $t=70$.}
\label{disc_surfacedenadia}
\end{figure}
In Fig.~\ref{countsigma}, we plot the surface densities of all runs (averaging over orbits $45-90$).
In the top panel we show the ${\tt retro}$ profiles as a function of radius in units of \a. 
As \ecc\, increases, the edge of the cavity is pushed outwards, and the profile of the disc inside the cavity becomes shallower. 
The lower panel shows the profiles of the four ${\tt count}$ setups in comparison with the prograde, circular setup,
as a function of radius now in units of \a(1+\ecc).
For the  ${\tt count}$ runs $\Sigma(r)$ are very similar and the radial displacement in $\Sigma_{\rm max}$ between prograde and retrograde 
is one semi--major axis (i.e., the disc extends inwards because of lack of resonances). In Fig.~\ref{disc_surfacedenadia}, the logarithmic surface density inside $5\a$ is shown for the two  runs ${\tt retro\,\ecc=0}$ and ${\tt retro\,\ecc=08}$, with color code and scale identical to paperI, 
Fig. 9. We can appreciate here the much smaller and well defined cavity in the circular case compared to the eccentric one.

The question of the necessary resolution is addressed both by changing the total number of SPH particles and by changing the mass of each SPH
particle (thus giving a different disc thickness, which must be sufficiently sampled)
As can be seen by inspecting table~\ref{initretro}, the ${\tt retro}$ runs use $2$, the ${\tt count}$ 
runs use $3$ and the ${\tt retro^{**}}$ runs use $4$ Million particles (where we state the resolution {\it after}
discarding relaxation). We measured differential torque--densities, \tordens (see \S 3.2), per 
unit mass in ${\tt retro\,\ecc=0}$ versus ${\tt count00}$, and found small, but no significant differences between the averaged torques
for the circular runs. This assures us that the resolution is sufficient in the ${\tt retro}$ runs, that the results are not 
sensitive to the precise choice of disc mass and that the relaxation of the discs to the new BHB orbit is fast (the ${\tt count 00}$ run 
effectively discards the first $795$ orbits of relaxation at which time it has a resolution of $3$ Million).
To be conservative, we use $4$ Million particles and monitor closely all conservation laws
 for the more complex dynamical geometry discussed in the end of the paper.
 
\section{Analysis of the numerical data}
\label{sec:analysis}

\subsection{Binary evolution: general features}
\begin{figure*}
\centering
\begin{tabular}{ccc}
\includegraphics[width=5.5cm,clip=true,angle=0]{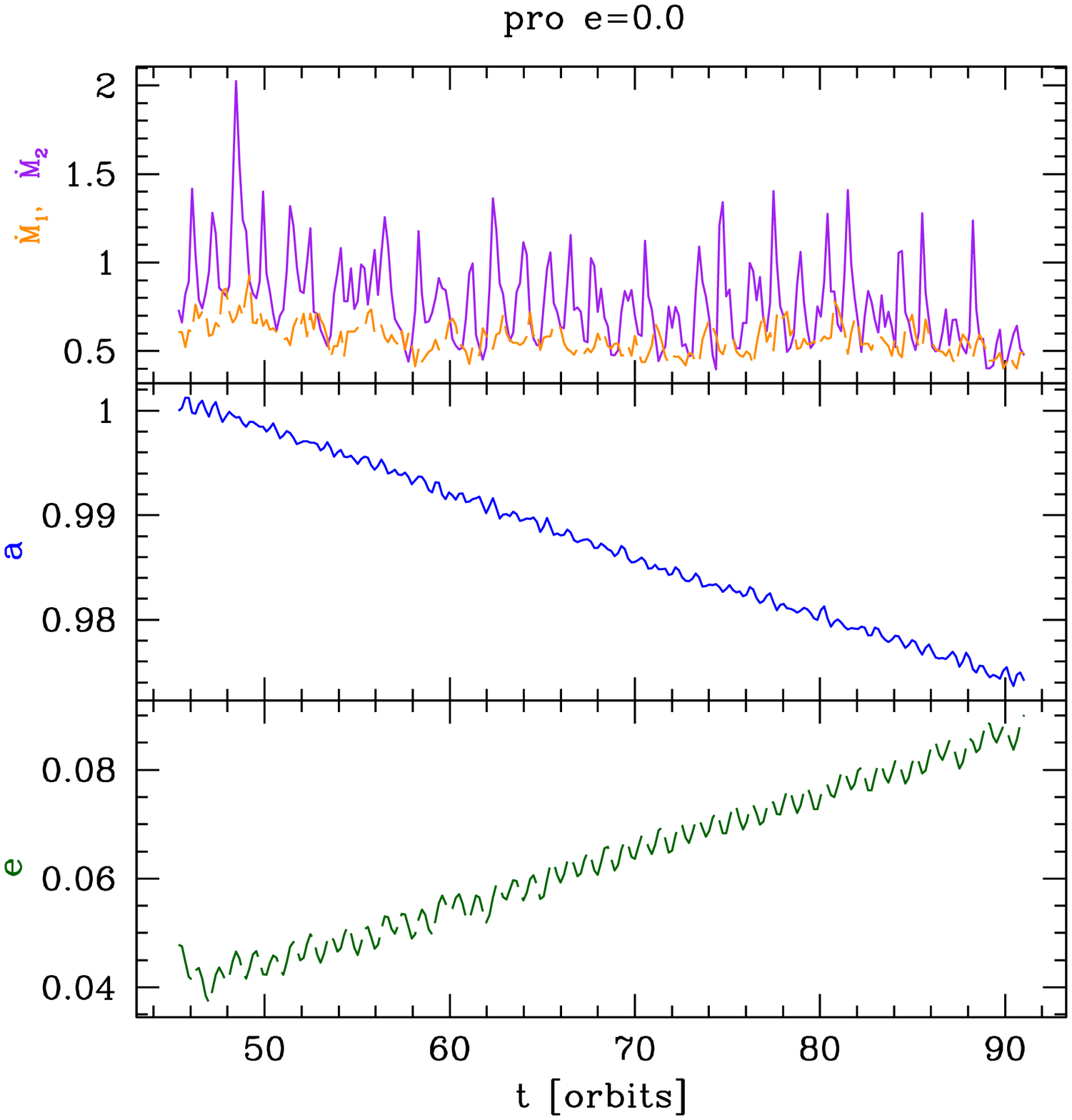}&
\includegraphics[width=5.5cm,clip=true,angle=0]{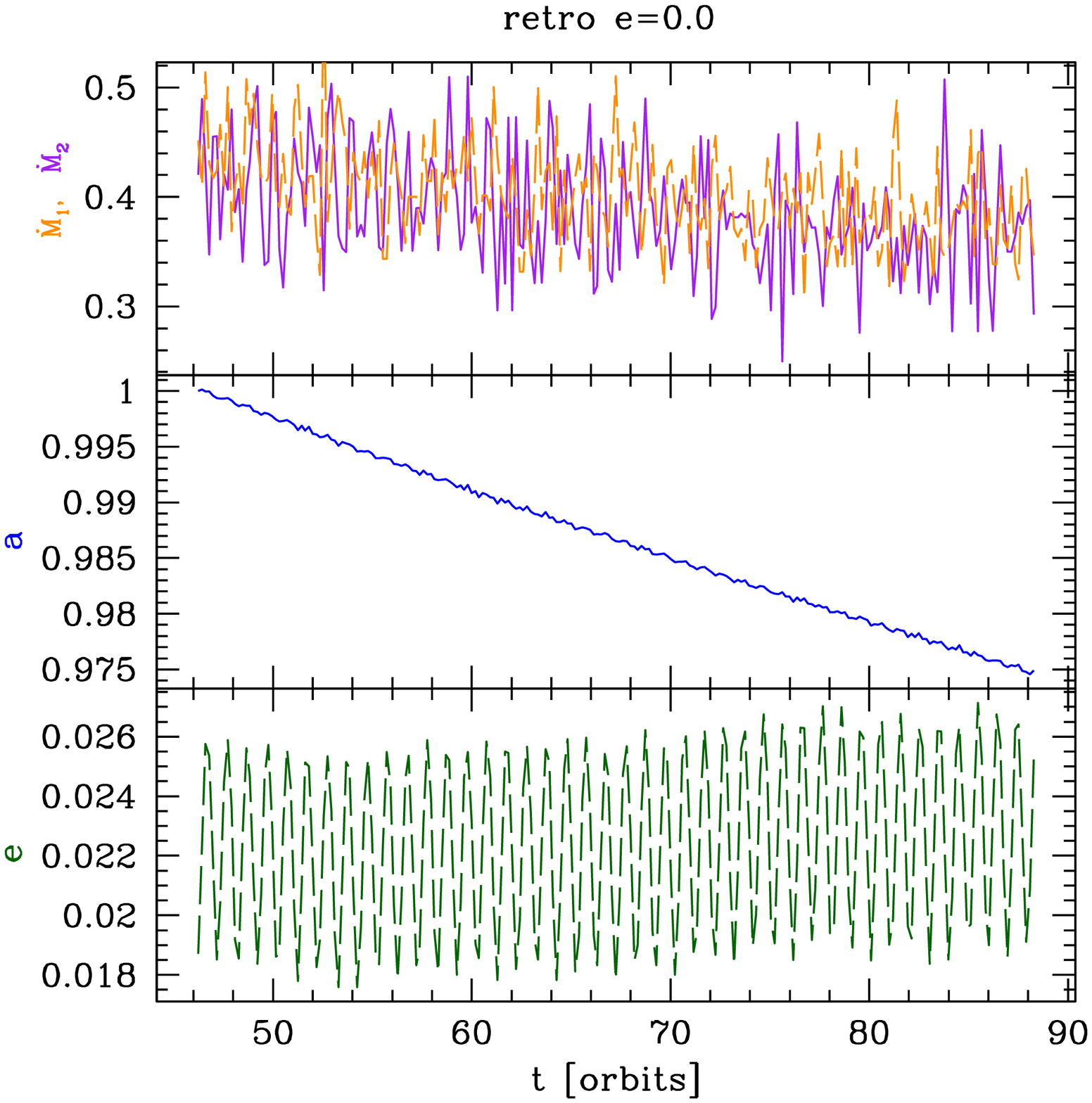}&
\includegraphics[width=5.5cm,clip=true,angle=0]{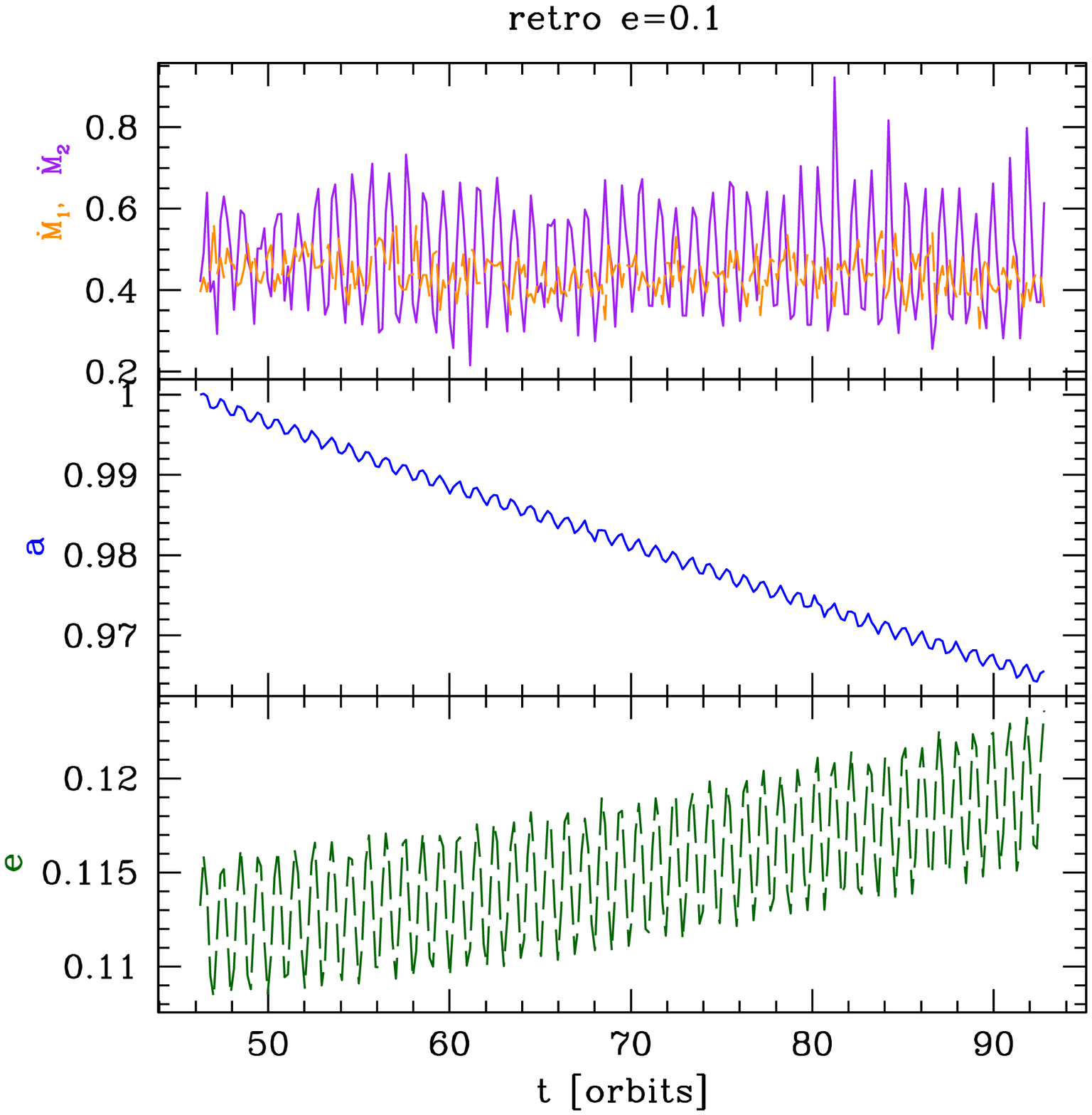}\\
\includegraphics[width=5.5cm,clip=true,angle=0]{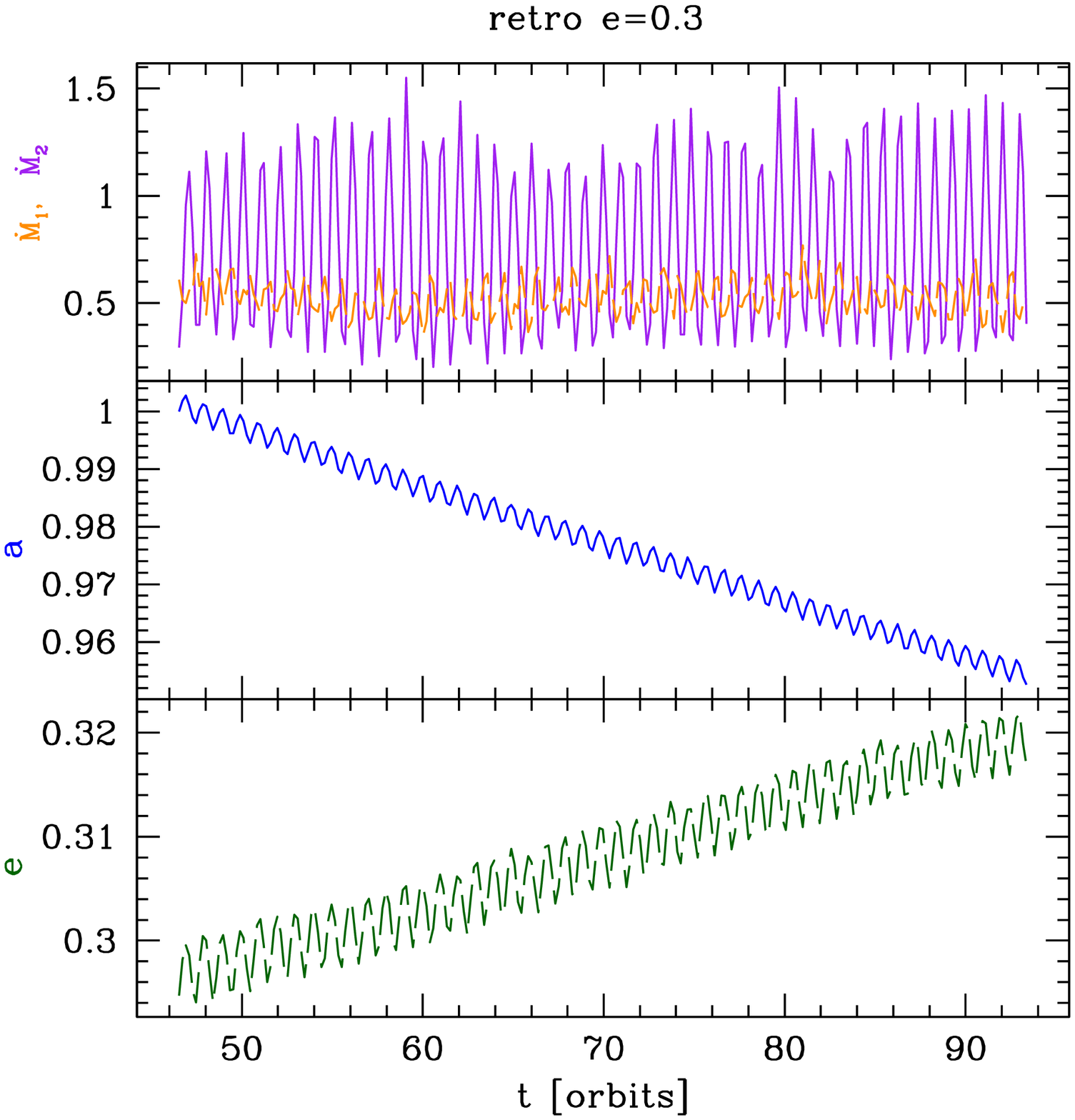}&
\includegraphics[width=5.5cm,clip=true,angle=0]{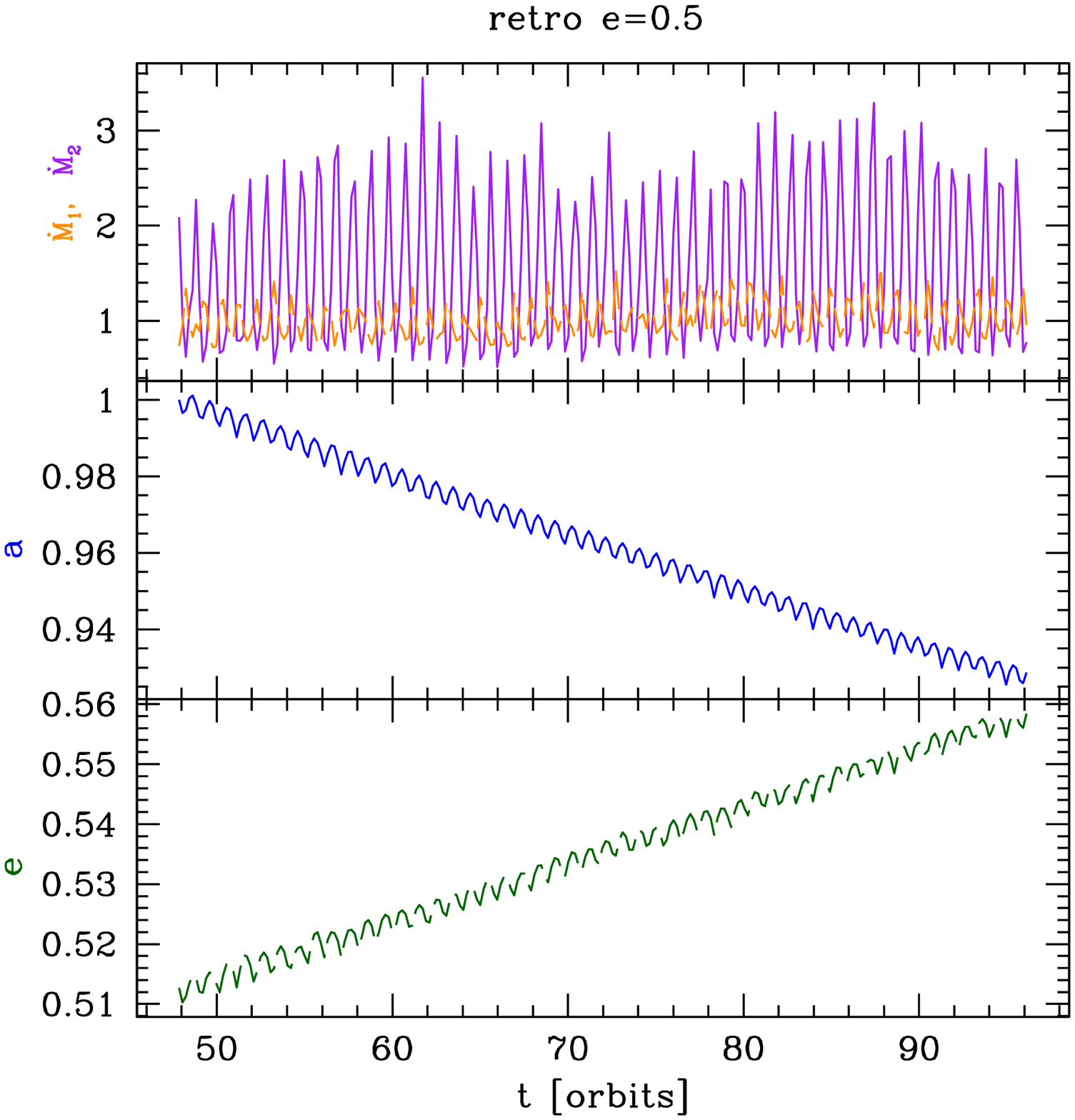}&
\includegraphics[width=5.5cm,clip=true,angle=0]{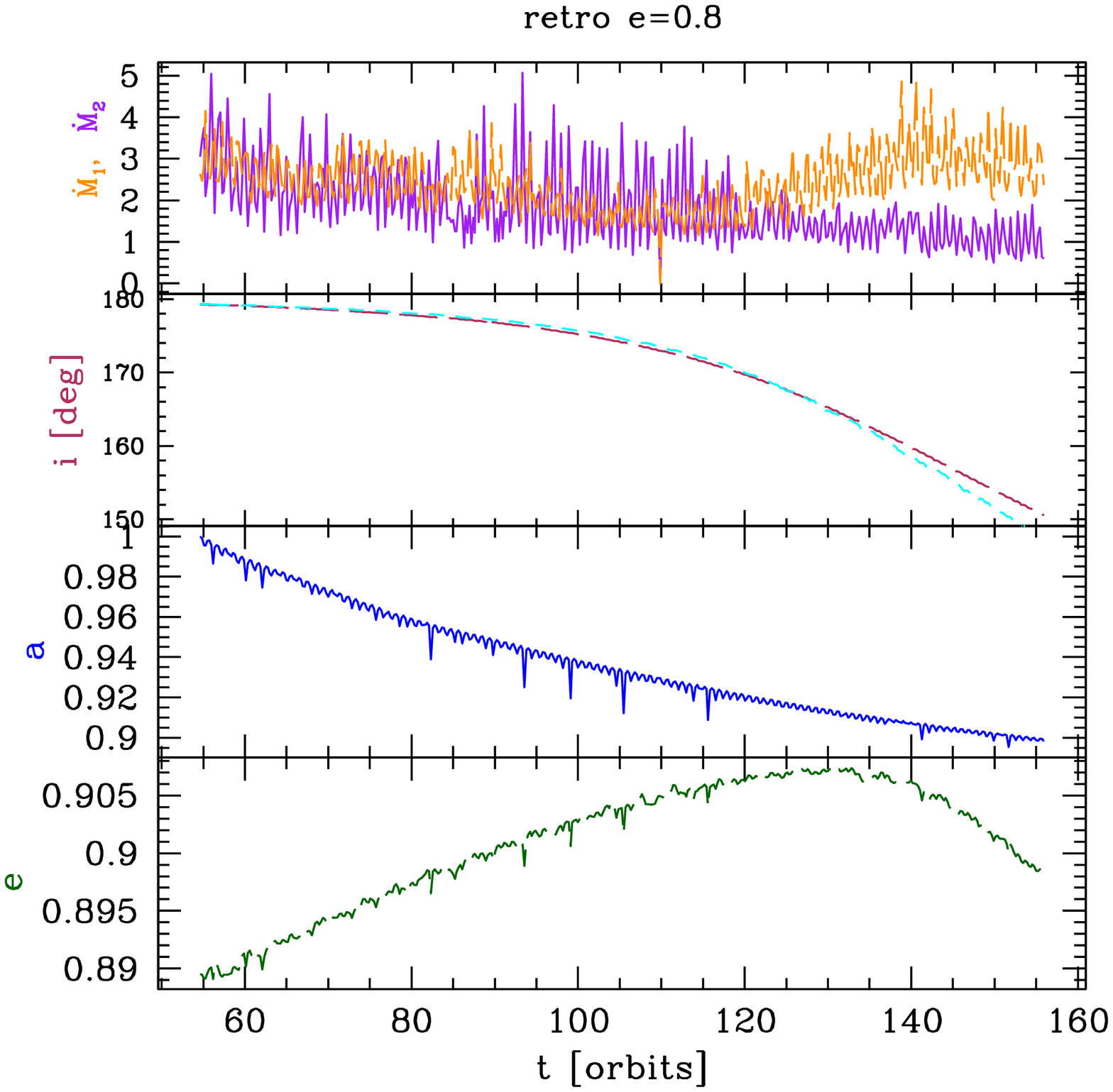}
\end{tabular}
\caption{{\sc Evolution of the accretion rates, semi-major axis, eccentricity and inclination} - All ${\tt retro}$ runs are shown, as indicated above each panel. We also show the prograde run in the top left panel. In each panel we plot the individual accretion rates \textcolor{orange}{$\dot{M}_1$}, \textcolor{RoyalPurple}{$\dot{M}_2$} (multiplied by $10^5$), the  \textcolor{blue}{semi-major axis $\a$} and the \textcolor{OliveGreen}{eccentricity $\ecc$}. In the ${\tt retro\,\ecc= 0.8}$ panel we also include the relative BHB-disc \textcolor{Maroon}{inclination $\i$}
in degrees. Note the different scales on the $y$ axis; quantities are in code units.}
\label{countlow}
\end{figure*}
We start with a qualitative description of the several outputs of the simulations. In Fig.~\ref{countlow}, we show the measured evolution of the individual accretion rates \textcolor{orange}{$\dot{M}_1$}, \textcolor{RoyalPurple}{$\dot{M}_2$}, \textcolor{blue}{semi-major axis $\a$} (for clarity normalized to \a(t=$45$)) and the \textcolor{OliveGreen}{eccentricity $\ecc$} for all the ${\tt retro}$ runs. The circular prograde case is shown in the top left panel for comparison. For the highest \ecc, we add a panel showing the
\textcolor{Maroon}{inclination $\i$}\footnote{Defined as 
$\i=\arccos\left(\frac{\langle {\bf L}_{\rm BHB}|{\bf L}_{\rm Disc}\rangle}{|{\bf L}_{\rm BHB}| |{\bf L}_{\rm Disc}|}\right)$} 
in degrees; $180$\textdegree\, implies perfect antialignment (an additional line shows the inclination \textcolor{cyan}{ i$_{2.5}$}
of only the material inside of $r<2.5\a(1+\ecc)$). In all cases $\langle\dot{\a}\rangle<0$ (here and in the following
$\langle\ast\rangle$ indicates time average) and its absolute value 
increases with eccentricity; i.e., more eccentric counterrotating binaries appear to shrink faster. 
$\langle\dot{\ecc}\rangle$ is positive for $\ecc>0$, but it is almost exactly zero for the quasi-circular run (${\tt retro}\,\, \ecc=0$).
This is in contrast to the circular prograde case, which features $\langle\dot{\ecc}\rangle>0$ (top left panel). 
The accretion rate onto each hole is comparable in the ${\tt retro}\,\, \ecc=0$ and it is smaller than the prograde case.
With increasing \ecc,  the accretion onto the secondary hole becomes larger and highly periodic (modulated on the orbital period of the binary).
The ${\tt retro}\,\, \ecc=0.8$ case shows a different general behavior. 
The binary leaves the disc orbital plane, making accretion on the primary hole easier (the 
secondary hole does not 'screen' the primary anymore) and the eccentricity growth has a turnover at \ecc=0.9.
We will return on this latter run extensively in \S \ref{sec:tilting}.
    
\subsection{Gravitational and accretion torques}
\label{sec:gravtorx}
\begin{figure*}
\centering
\includegraphics[width=0.95\linewidth,trim= 0cm 1cm 0cm 5.5cm,clip]{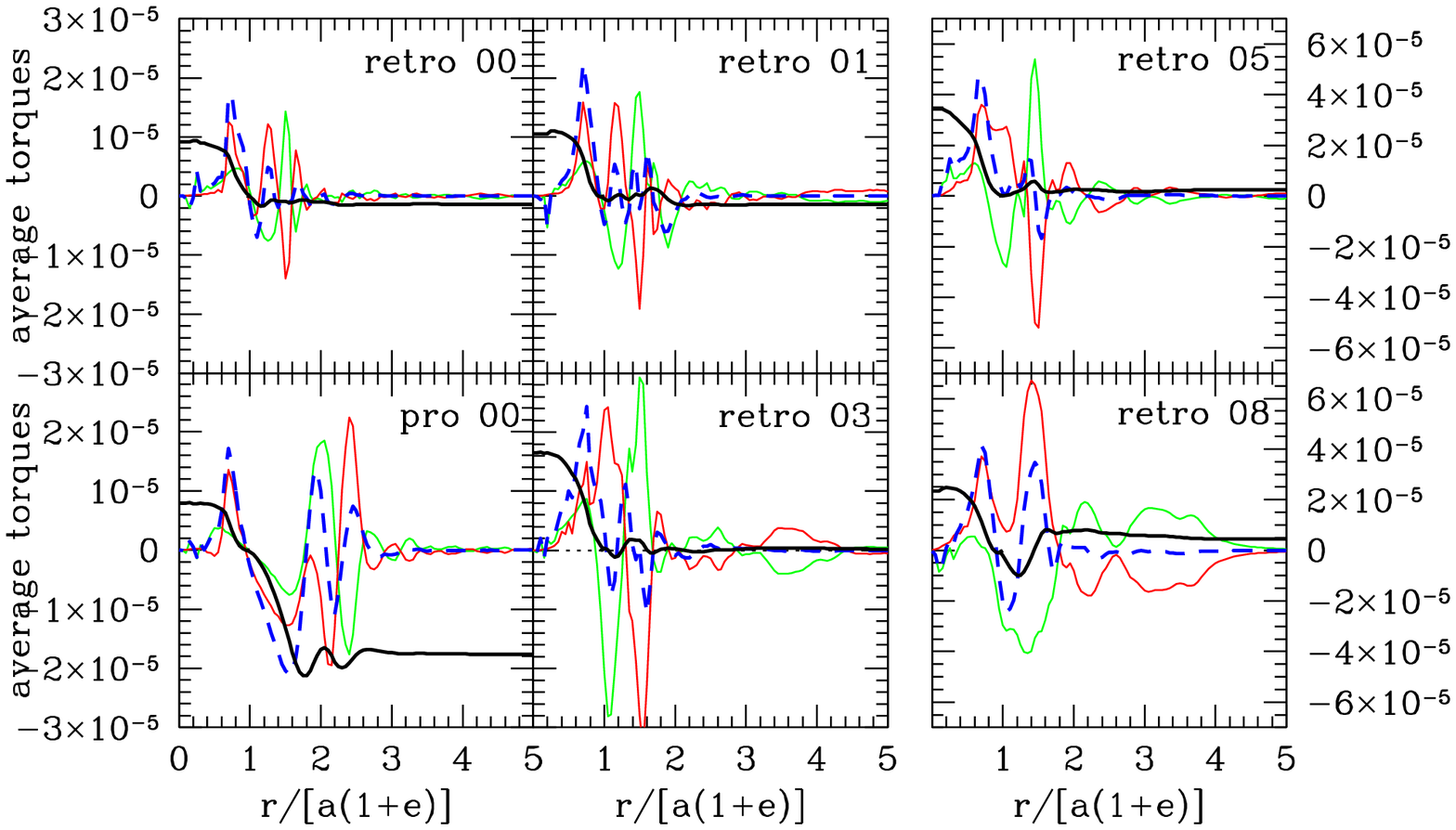}
\caption{
{\bf \sc Gravitational torques as a function of the distance to the BHB CoM} -
In each panel the differential torques \tordens
exerted by the disc onto the BHs are averaged over the orbits $45-90$ and shown in units of $[\G M^2_0(\a(1+\ecc))^{-2}]$. We show the differential torque on the primary (\textcolor{red}{red}), on the secondary  (\textcolor{green}{green}), the sum of the two (\textcolor{blue}{dashed--blue}), and the total torques integrated {\it inwards} and {\it outwards} (\textcolor{black}{black}). Note that the angular momentum of the BHB is negative for retrograde runs and positive for prograde runs. See text for details.}
\label{torquecompare}
\end{figure*}
As elucidated in paperI the evolution of the binary elements is driven by the sum of the gravitational plus the accretion torques: ${\bf \dot{L}}_{\rm BHB}={\bf T}_{\rm G}+{\bf T}_{\rm acc}${\footnote{Throughout the paper we use boldface for vectors.}}.
The total gravitational torque exerted by the sum of the gaseous particles on each individual BH can be computed as
\begin{equation}
{\bf T}_{\rm G}=\sum_{j=1}^N\sum_{k=1}^2 {\bf r}_k\times\frac{\G M_km_j({\bf r}_j-{\bf r}_k)}{|{\bf r}_j-{\bf r}_k|^3},
\label{torque}
\end{equation}
where ${\bf r}$ is the position vector with respect to the system CoM. The accretion torque ${\bf T}_{\rm acc}$ is measured from the conservation of linear momentum in the accretion process of each particle, i.e, ${\bf T}_{\rm acc}={\bf r}_k\times m_j{\bf v}_j$ (here ${\bf r}_k$ is the position vector of the accreting BH, and  $m_j$ and ${\bf v}_j$ are the mass and velocity of the accreted gas particle). In the following, all torques are averaged both in time and azimuth.
In all ${\tt retro}$ runs, we use a reference frame in which the angular momentum of the disc, ${\bf L_{\rm disc}}$, is initially aligned to the $z$ axis and points in the positive $z$-direction, whereas ${\bf L_{\rm BHB}}$ points in the negative $z$-direction. Assuming that the binary and the disc are coplanar, the system is effective two-dimensional from the dynamical standpoint. We can therefore safely take ${\bf L}\approx L_z$ and ${\bf T}\approx T_z$, and unless otherwise specified, we will always refer to the $z$ components of these quantities{\footnote{This does 
not apply to the  ${\tt retro}\,\, \ecc=0.8$ case that will be discussed separately in \S \ref{sec:tilting}.}}. 

\begin{figure*}
\centering
\includegraphics[width=0.95\linewidth,trim= 0cm 0cm 0cm 0cm,clip]{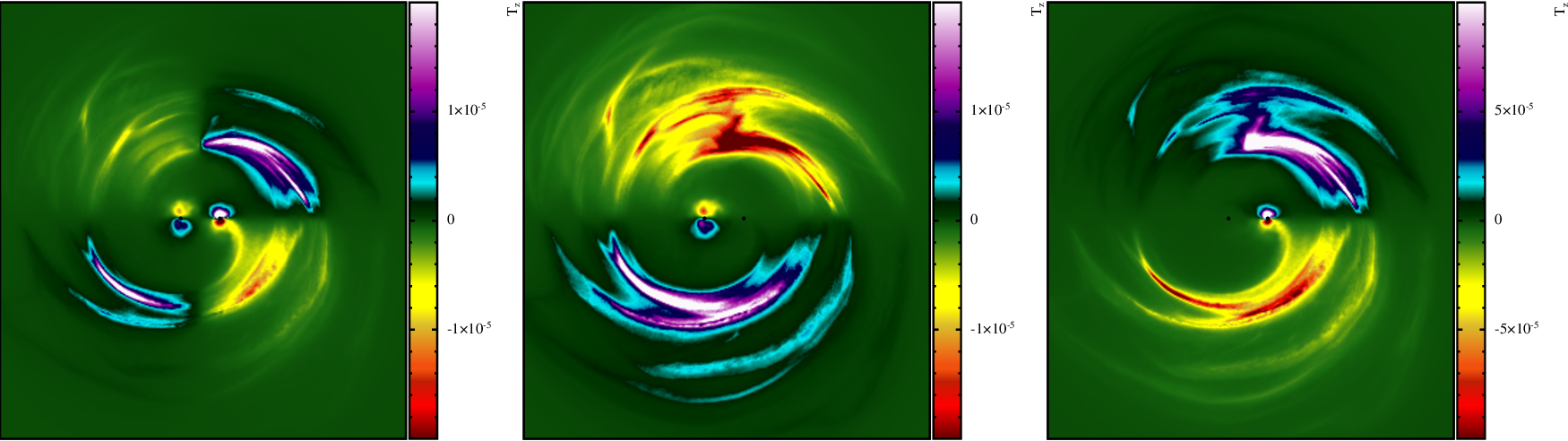}
\includegraphics[width=0.95\linewidth,trim= 0cm 0cm 0cm 0cm,clip]{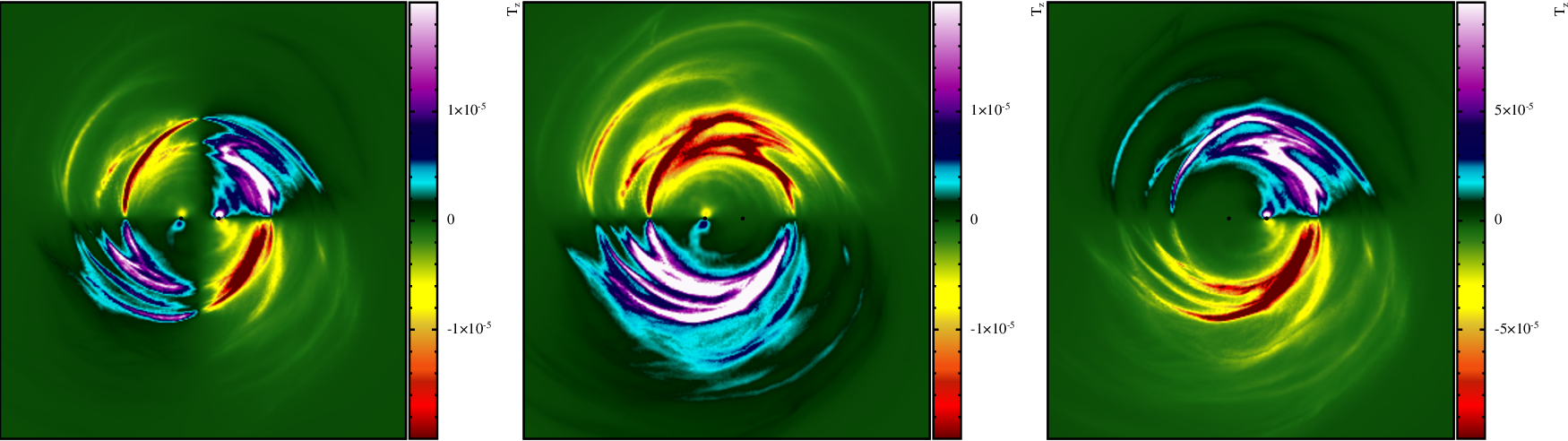}
\caption{
{\bf \sc Gravitational torque density} - Differential torque density in the $z$-direction 
for two initially circular BHB runs after $70$ orbits. 
Top (bottom) panels: ${\tt pro\,\ecc=0}$ (${\tt retro\,\ecc=0}$). In each row, from left to right we show the total torque
density on the binary  $T_{{\rm G},z}= M_1 \, T_{{\rm G},z_1}+ M_2 \, T_{{\rm G},z_2}$, the torque onto the primary BH, $T_{{\rm G},z_1}$,
and the torque onto the secondary BH, $ T_{{\rm G},z_2}$. Figure made using {\sc SPLASH} \citep{Price2007}.}
\label{torquedensity}
\end{figure*}

As discussed in the previous section, the natural scale of the system is the apocenter distance of the BHB. We thus integrate the
differential torques both in-- and outwards of $\a(1+\ecc)$ as:
\begin{equation}
\langle T_{z,[a,b]}\rangle=\left\langle\int_a^b\frac{dT_z}{dr}dr\right\rangle,
\label{eqavtorque}   
\end{equation}
In Fig.~\ref{torquecompare}, we show the differential, gravitational torque \tordens on the primary BH (\textcolor{red}{red}), 
 on the secondary BH (\textcolor{green}{green}), the sum of the two (\textcolor{blue}{blue}), and the total integrated torque 
(\textcolor{black}{black}). In opposition to the prograde case, the figure clarifies that the interaction between the disc
and the binary takes place mostly at the location of the secondary, supporting a direct 'BH-gas impact' model, as described by
\cite{nixon11a}. The differential torque density, \tordens, of the runs ${\tt pro\,\ecc=0}$ and ${\tt retro\,\ecc=0}$ is shown 
in Fig.~\ref{torquedensity}, where the leftmost panels show the total \tordens on the binary, the middle panel the 
torque density onto the primary BH and the right, onto the secondary BH.
Here, we can clearly appreciate the absence of a gap between \a\, and 2\a\,
in the retrograde run. 
The lower left panel highlights the unbalanced positive torque (\textcolor{blue}{blue})
at the location of the secondary hole, which is responsible for the binary evolution
(Note, that the orbital angular momentum of the binary is negative in the $\tt 
retro$ runs). 

Differentiating ${\bf L_{\rm BHB}}$  with respect to the relevant binary elements, angular momentum conservation yields:
\begin{equation}
\frac{\dot{\a}}{\a}=\textcolor{MidnightBlue}{\frac{2\,T_{\rm G_z} }{L_z}}+
\textcolor{Maroon}{\frac{2\,T_{\rm acc_z} }{L_z}}-
\frac{\dot{\textcolor{red}{M}}}{\textcolor{red}{M}}-\frac{\textcolor{Orange}{2}\dot{\mmu}}{\mmu}
+\textcolor{OliveGreen}{\frac{2\ecc}{1-\ecc^2}\dot{\ecc}},
\label{adotdec}   
\end{equation}

where $\mmu=M_1M_2/M$ is the binary reduced mass. Note that all quantities in Eq.~(\ref{adotdec})
can be directly measured from the simulation outputs. We can therefore use it to get an empirical measurement 
on how the angular momentum exchange is redistributed among the relevant binary elements. We have cast Eq.~(\ref{adotdec})
in terms of the shrinkage rate and 
 show in Fig.~\ref{countdelta} that the two major contributions are $\textcolor{MidnightBlue}{\frac{2\,T_{\rm G_z} }{L_z}}$
 and $\textcolor{OliveGreen}{\frac{2\ecc}{1-\ecc^2}\dot{\ecc}}$ which are of opposite sign and intrinsically coupled with each other.
From this decomposition (cast in terms of the orbital element of choice) 
we can get a heuristic understanding of the BHB-disc coupling, which will serve as a guide for an analytical
interpretation of the results.

\subsection{Comparison of circular runs: prograde vs retrograde phenomenology}
\label{sec:retroI}
\begin{figure}
\centering
  \includegraphics[width=0.95\linewidth]{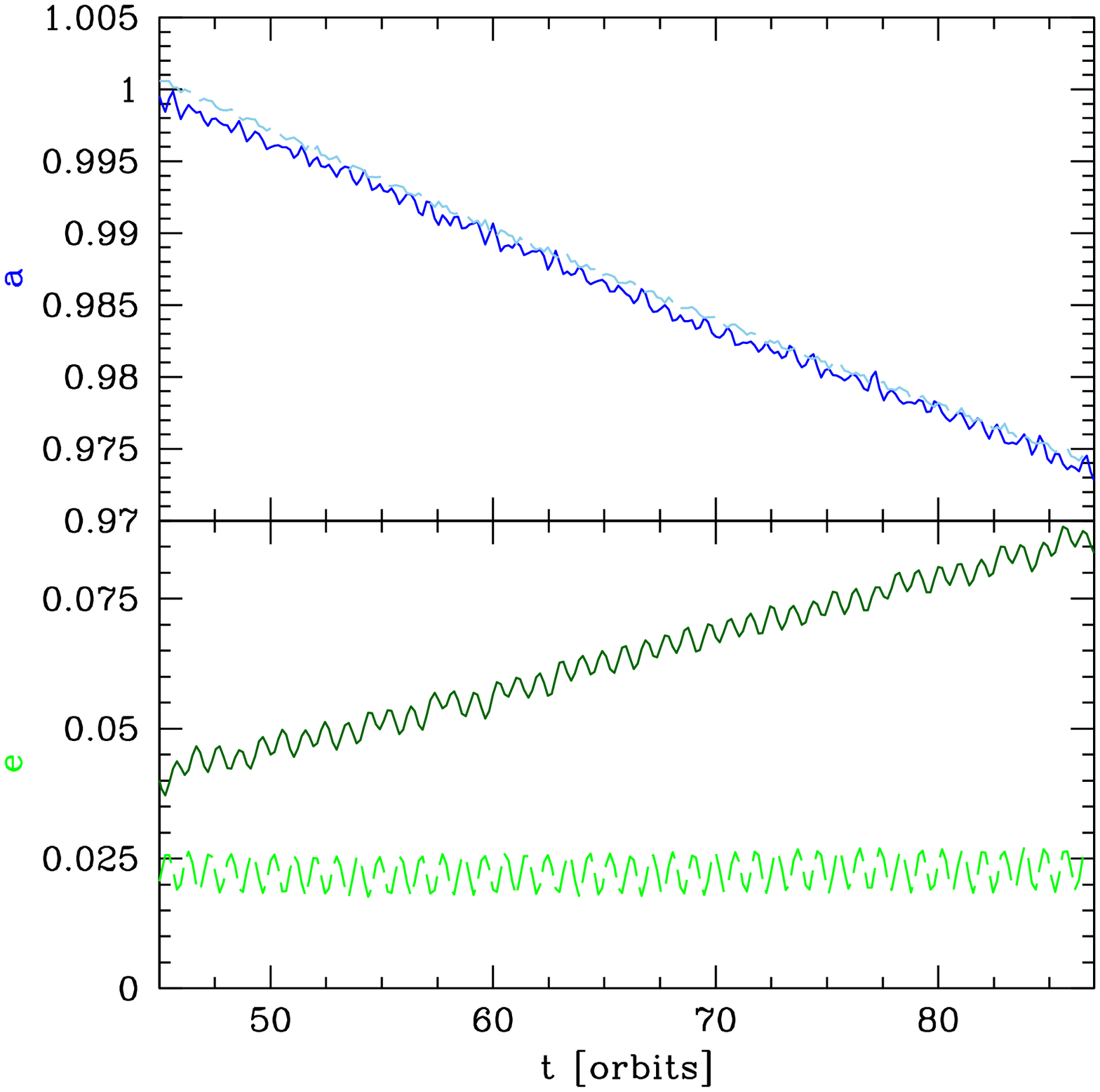}
\caption{{\sc Evolution of semi major axis and eccentricity for circular BHBs } -
Direct comparison of the prograde (solid lines) and retrograde (dashed lines) systems. 
Only the stage after the initial transient is shown. The evolution of \a\, cannot be distinguished between the two setups,
whereas the evolution of \ecc\, is substantially different.
}
\label{eccsemicompare}
\end{figure}

As we chose the setup of the  ${\tt retro}$ runs to be the same as those of the  {\it adia05} run of paperI, we can compare the evolution of \a\, and \ecc\, directly. In Fig.~\ref{eccsemicompare}, we thus plot the circular runs $\mathtt{retro\, \ecc=0}$ vs ${\tt pro\,\ecc=0}$. The evolution of \a\, is not distinguishable over the $45$ orbits that we show; on the other hand, the eccentricity behavior is different, such that $\dot{\ecc}|_{\rm retro}=0$ whereas  $\dot{\ecc}|_{\rm pro}>0$. Upon inspecting Fig.~\ref{countlow}, we find that also the accretion rates for the prograde setup  are larger and that while the individual rates are equal for the retrograde, the secondary BH accretes and varies more in the prograde.
Considering now the very left panels of Fig.~\ref{torquecompare}, focusing only on the gravitational 
{\it inwards} torque, $T_{{\rm in},z}$, we find them to be very similar, but opposite in sign with respect to $L_{{\rm BHB},z}$.
However, the integrated $T_{{\rm in},z}$ is slightly smaller in the prograde and there are three clear differences: 
\begin{enumerate}
\item the radial location of largest (differential) torque amplitude onto the individual holes are displaced by a factor  \a\,;
\item the largest peak of the (differential) torque onto the binary (blue dashed line) is around $r\sim 1.7 \a(1+\ecc)$,
(coinciding with the first OLR)
where the torques onto the individual holes are in phase, is absent in the retrograde case;
\item the substantial {\it outwards} torques found in the prograde case are completely absent in the retrograde case; 
\end{enumerate}

\begin{figure}
\centering
\includegraphics[width=7.2cm,clip=true]{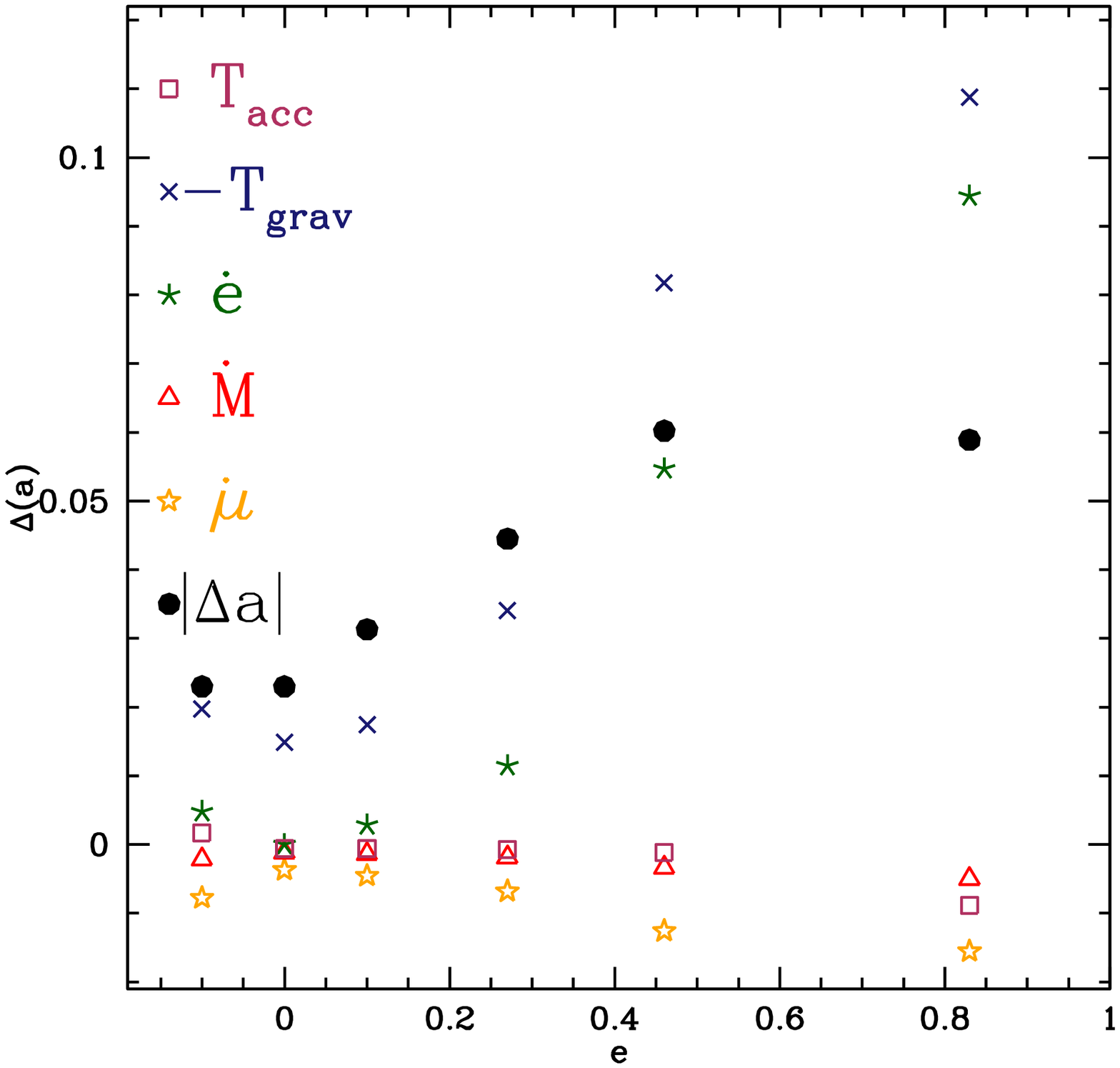}
\caption{{\sc Contributions to $\dot{\a}$ by individual components}; the ${\tt pro 00}$ run is inserted as negative \ecc. 
Terms color--coded according to Eq.~\ref{adotdec}.
}
\label{countdelta}
\end{figure}

Note that, integrated across the whole $r$ range, the total torques on the binary always have opposite sign with respect to the orientation
of {\bf L}$_{\rm BHB}$, causing a net transfer of angular momentum from the binary to the disc {\it both in the prograde and in the retrograde case}. 
Turning to the redistribution of the angular momentum transfer among the binary elements as detailed by equation (\ref{adotdec}),
we show the individual terms from the perspective of $\dot{\a}$
in Fig.~\ref{countdelta} and first compare the two leftmost (i.e. circular) runs\footnote{The prograde
run is inserted at negative \ecc, just to put it next to its retrograde counterpart}.
We observe that
\begin{enumerate}
\item the contribution of $\textcolor{OliveGreen}{\frac{2\ecc}{1-\ecc^2}\dot{\ecc}}$ (\textcolor{OliveGreen}{filled pentagons}) is negligible for the retrograde as $\dot{\ecc}=0$
\item\textcolor{Maroon}{ $2\, T_{{\rm acc},z}/L_z$} (\textcolor{Maroon}{open squares}) contributes negatively for the retrograde, whereas positively (and three times stronger) for the prograde. The sign simply reflects the sign of $L_z$. 
 \item the mass transfer factors $\textcolor{Orange}{2}\,\dot{\mmu}/\mmu$ and $\textcolor{red}{\dot{M}/M}$ are larger by a factor $\sim$ 2 in the prograde case.
 \item summing up all components gives a very similar $\dot{\a}$ 
\end{enumerate}
Taken together, these points show that for the retrograde case, the gravitational torque is indeed directly related to the binary semi--major axis evolution, given that the distribution of angular momentum among the other elements of the binary (mass, mass ratio and eccentricity) is negligible. 

\subsection{Eccentric runs}
\label{sec:ecc}
We turn now to the description of the phenomenology of eccentric runs. In the following, 
we provide some analytical fits to the few measured data points provided by our set of 5 simulations 
(\{${\tt retro\,\ecc=0\,\ecc=0.1\,\ecc=0.3\,\ecc=0.5\, \ecc=0.8}$\}). 
We note that sometimes either the circular or the tilting run (${\tt\ecc=0.8}$) had to be excluded, and we caution that the fitting functions have only few degree of freedom less than the number of sample points, which is why we do not pursue any extrapolation or other use of the numerical fit-coefficients.

\begin{figure}
\centering
\includegraphics[width=7.2cm,clip=true]{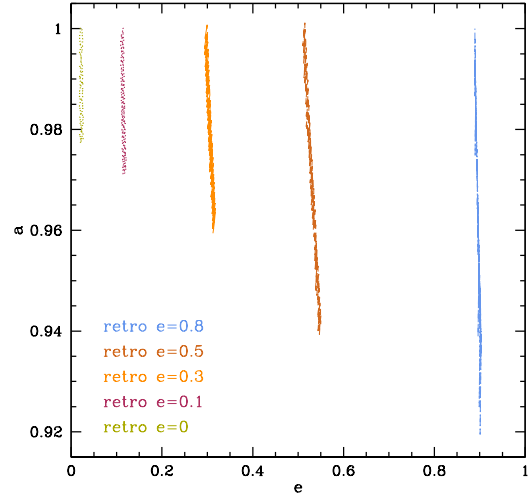}
\caption{{\sc Semi--major axis \a\,,  and eccentricity \ecc\, evolution} in the {\tt retro} runs. All runs cover the same time interval. See text for details.
}
\label{countdelta_ae}
\end{figure}

In Fig.~\ref{torquecompare}, we plot the evolution of the torque structure from circular to eccentric,
and in Fig.~\ref{countdelta} the contributions to the shrinkage rate, $\dot{\a}$ are given,
 also using the total integrated torque components. 
 Clear trends with increasing \ecc\, are:
\begin{enumerate}
 \item with the exception of the run ${\tt retro\, \ecc=0.8}$, the integrated {\it inward} torque grows  with \ecc\, as 
 \beq
 T_{{\rm in},z}(\ecc)\approx T_{{\rm in},z}(\ecc=0)+ 10\cdot T_{{\rm in},z}(\ecc=0)\, \ecc^2;
 \eeq
 \item  the integrated {\it outward} torque grows linearly with \ecc\,, for $\ecc\neq0$, but is always negligible with respect to the {\it inward} torque;
 \item a positive {\it bump} appears in the {\it outward} torque around $r\sim 1.5 \a(1+\ecc)$, for $\ecc\neq0$;
 \item the  amplitude of the differential torques, \tordens, grows with \ecc;
 \item the  maximum radius outside of which \tordens $\sim 0$  is approximately  $r_{\rm max}\lsim 2.5 \a(1+\ecc)$ (cf. dashed blue lines).
 \item ${\tt retro\, \ecc=0.8}$ shows a qualitatively different torque structure,
 namely there is a large differential torque outside of $r=\a(1+\ecc)$,
 there is a high and positive net torque resulting from the fact, 
 that now {\it pairs of two} subsequent individual, differential torques (either red or green) alternate in sign,
 rather than {\it single} peaks alternating after every extreme (all other panels).
 We conjecture, that this reflects  the BHB leaving the plane of the disc;
 an effect that will be discussed further in \S~\ref{sec:tilting}.
\end{enumerate}
Measuring the derivatives of the orbital elements as a function of \ecc, we additionally find
(cf. Figs~\ref{countdelta_ae} and~\ref{countmm}):
\begin{itemize} 
\item the dependence of $\dot{\ecc}(\ecc)$ (as long as the inclination stays zero) is linear and positive:
\beq
\dot{\ecc}(\ecc)\approx -0.0034+0.09\,\ecc
\label{edot}
\eeq
implying a threshold eccentricity of $\ecc\sim 0.04$, for where the growth of \ecc\, would be excited;
\item the dependence of $\dot{\a}(\ecc)$ is linear and negative;
\item the dependence of $\mbhbdot(\ecc)$ is linear and positive;
\item the dependence of $\ratio(\ecc)$ is a concave parabola with a best fit maximum around $\ecc\sim 0.43$;
\end{itemize}

\subsection{Accretion rates}
\begin{figure}
\centering
\includegraphics[width=7.2cm,clip=true]{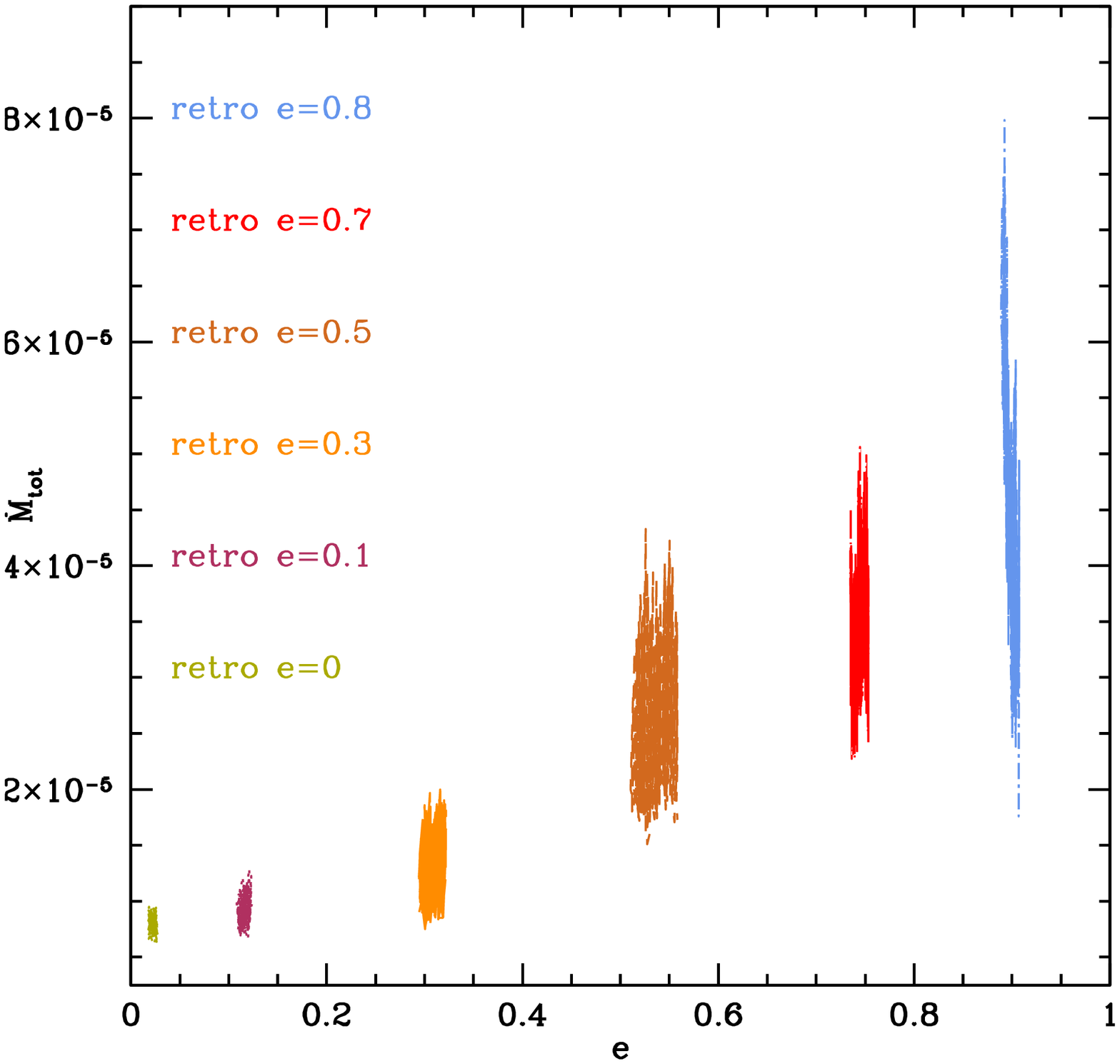}
\includegraphics[width=7.2cm,clip=true]{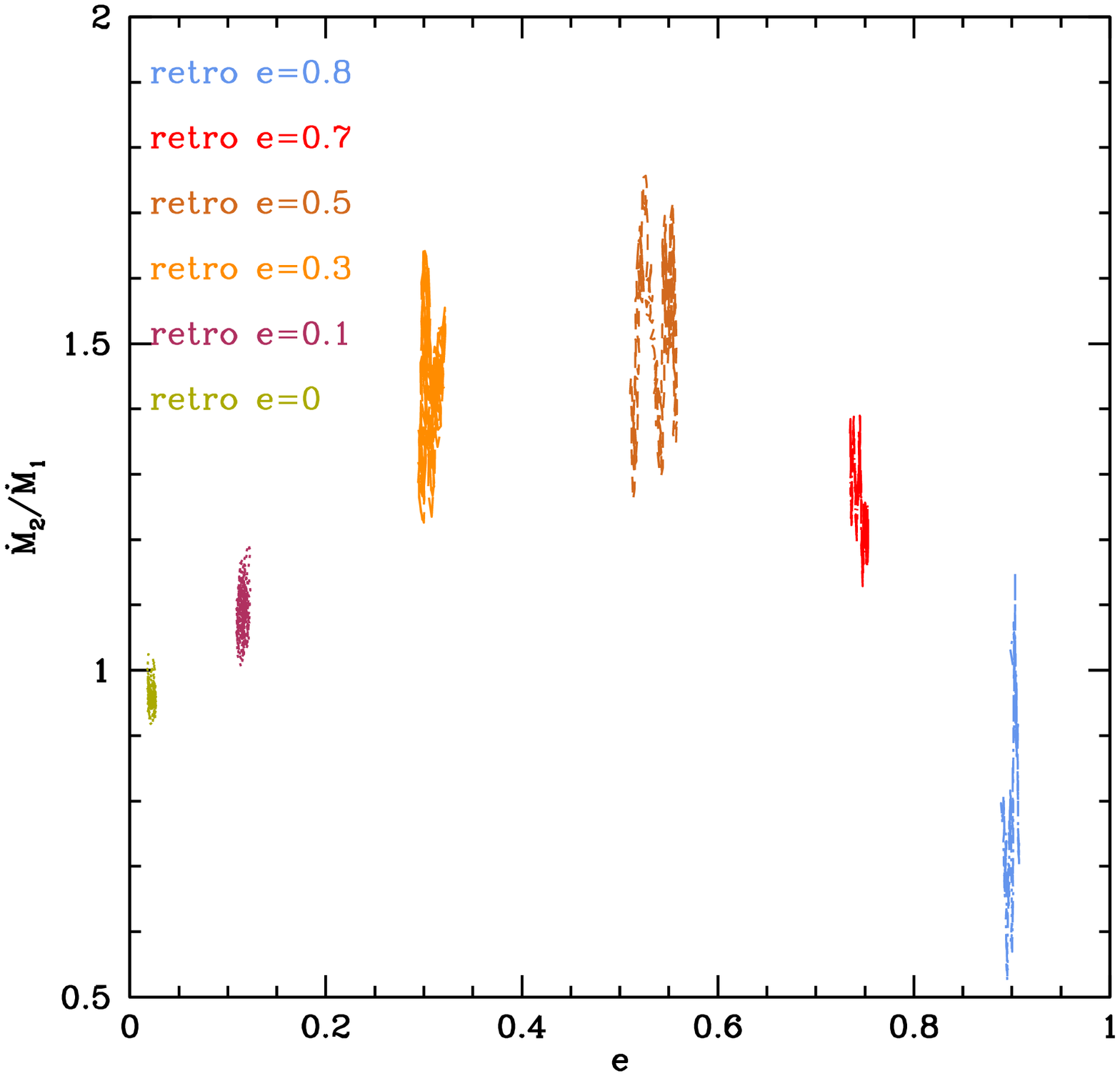}
\caption{
{\bf \sc Mass transfer rates as function of eccentricity } - 
({\it Top}): Total accretion rate for the different runs.  ({\it Bottom}): ratio between secondary  
and primary accretion rate $\dot{M_2}/\dot{M_1}$, as a function of eccentricity  for the retrograde runs.
All ${\tt retro}$ runs are plotted over the same time interval (orbits $45-90$).}
\label{countmm}
\end{figure}

In Fig.~\ref{countlow}, we can spot clear trends in both $\mbhbdot$ and $\ratio$ with $\ecc$. We explore this further in Fig.~\ref{countmm}, where we show the accretion rates, both their sum and their ratio, as a function of eccentricity. The combined accretion rate increases with increasing $\ecc$, in contrast to what was observed for prograde discs in paperI. The explanation is straightforward: for prograde runs the cavity is forced to expand with higher $\ecc$ due to resonances and slingshots whereas this is not the case for retrograde runs where instead the binary penetrates further and into the disc edge and thus accretes more. Note also that the accretion rate growth is proportional to the enhancement of the inwards torque in Fig. \ref{torquecompare} (with the exception of the ${\tt retro\,\ecc=0.8}$ case), establishing a clear link between the two quantities. The ratio of secondary to primary accretion rate reaches a maximum somewhere in the range $0.4<\ecc<0.6$ and declines for higher $\ecc$.
 Conversely, in the prograde case \citep{Roedig2011}, the ratio decreases to around $\ratio\sim 1$ only for very high $\ecc\sim 0.8$ and there is no turnover.

\section{Analytical modeling}
\label{sec:analytics}
We now show that the phenomenology described in the previous section can be interpreted in the light of a simple 'dust model' 
in which particles experience a pure gravitational interaction with the BHB . 

\subsection{Retrograde runs}
The retrograde case has been already described by \citet[hereafter N11,][]{nixon11a}. 
Following the same argument, we assume that a pure kinetic interaction between the secondary
BH and the gas occurs at each apoastron passage. Since in our case $\q\approx 1/3$, we relax the assumption $\q\ll1$,
but we keep the Ansatz that the inner boundary of the disc interacts with the secondary hole only.
A close inspection of the disc-BHB dynamics shows that the gas that 'impacts' on the secondary hole is partly 
accreted and partly slowed down to the point that is easily captured by the primary. 
We therefore assume that when a mass $\Delta{M}$ interacts with the secondary, a fraction $0<\alpha<1$ 
is accreted by $\mzwei$, and the remaining $1-\alpha$ is captured by $\meins$.
We can therefore impose conservation of linear momentum in the interaction to get (see N11 for the mathematical derivation)
\begin{equation}
\frac{\Delta \a}{\a}=-\frac{2\Delta{M}}{M_2(1+\ecc)}{\cal F}(\ecc,\q,\alpha),
\label{deltaa}
\end{equation}
\begin{equation}
{\Delta \ecc}=\frac{2\Delta{M}}{M_2}{\cal F}(\ecc,\q,\alpha),
\label{deltae}
\end{equation}
where 
\begin{equation}
{\cal F}(\ecc,\q,\alpha)=(1-\ecc)^{1/2}(1+\q)^{1/2}+\alpha\frac{1-\ecc}{1+\q},
\label{deltae}
\end{equation}
and the $1+\q$ factors (absent in N11) stem from the fact that we relaxed the condition $\q\ll 1$. 
If we now average over the binary orbit, and we take the continuum limit,
the above expressions result in two coupled differential equations for $\dot{\a}$ and $\dot{\ecc}$,
as a function of the accretion rate $\dot{M}$ and the parameter $\alpha$.
A complete solution to the dynamics requires an additional knowledge of $\dot{M}$ and $\alpha$ as a function of $\a$ and $\ecc$. 
However Fig. \ref{countlow} indicates that on the short period of the simulations those are basically constant.
We also note that $\ecc$ enters in the RHS of the equations only through the combinations $1-\ecc$ and $1+\ecc$,
providing, over small timescales, only small corrections that we neglect.
We can therefore set $\ecc=\ecc_0$ in the RHS of equations (\ref{deltaa}) and (\ref{deltae}),
perform the integral via separation of variables and linearize the solution to get:
\begin{equation}
\a_t=\a_0\left[1-\frac{2\dot{M}}{M_2(1+\ecc_0)}{\cal F}(\ecc,\q,\alpha)t\right]
\end{equation}
\begin{equation}
\ecc_t=\ecc_0\left[1+\frac{2\Delta{M}}{M_2}{\cal F}(\ecc,\q,\alpha)t\right].
\end{equation}
The results are shown in table \ref{tabres} for $t=2\pi\times 45$ (this corresponds to $t=90$ in Fig. \ref{countlow}). 
A visual comparison with Fig. \ref{countlow} indicates that, although there are differences, 
the overall quantitative agreement is quite good. Discarding the run ${\tt retro\, \ecc=0.8}$ 
(the BHB leaves the plane and our analysis does not apply),
we notice however that the model slightly underestimate both the binary shrinking and the eccentricity growth for moderate $\ecc$
(runs ${\tt retro\, \ecc=0.3 \,, \ecc=0.5}$). 
\begin{table}
\caption{
{\bf \sc Analytical model: linearized evolution of the BHB}.
All ${\tt retro}$ runs are shown. $\a_{45}$ and $\ecc_{45}$ indicate the BHB semi--major axis 
and eccentricity after 45 orbits of binary evolution, this corresponds to $t=90$ in Fig. \ref{countlow}. 
\label{tabres}}
\begin{center}
\begin{tabular}{lcccccc}
\hline 
\hline
Model \quad& $\dot{M}(\times10^{-5})$  \quad& $\alpha$  \quad& $\a_0$  \quad& $\a_{45}$  \quad&  $\ecc_0$    \quad&  $\ecc_{45}$\quad\\
\hline
$\mathtt{retro\, \ecc=0}$    \quad& 0.8 & 0.50 & $1.0$ & 0.973 & $0.022$& 0.022\\
$\mathtt{retro\, \ecc=0.1}$  \quad& 1.0 & 0.52 & $1.0$ & 0.970 & $0.112$& 0.116\\
$\mathtt{retro\, \ecc=0.3}$  \quad& 1.4 & 0.58 & $1.0$ & 0.968 & $0.297$& 0.309\\
$\mathtt{retro\, \ecc=0.5}$  \quad& 2.8 & 0.60 & $1.0$ & 0.957 & $0.511$& 0.544\\
$\mathtt{retro\, \ecc=0.8}$  \quad& 5.5 & 0.44 & $1.0$ & 0.972 & $0.888$& 0.934\\
\hline
\end{tabular}
\end{center}
\end{table}
If we now approximate the torque exerted on the gas as $T_{\gas}=(dL/dt)_{\gas}\approx-\dot{M}\Delta{v}_{\gas}\a$ ($v_{\gas}$
refers to the gas velocity),
and consider that the gas is partly accreted by the secondary (i.e, $\Delta{v}_{\gas}\approx -2v_{M_{2}}$)
and partly slowed down to the point is accreted by $M_1$ (i.e., $\Delta{v}_{\gas}\approx -v_{M_{2}}$),
we obtain $T_{\gas}\approx\dot{M}$ (in simulation units), as observed in Fig. \ref{torquecompare}.
This also trivially explains the behavior of $T(\ecc)_{\rm total}$: for eccentric binaries,
the secondary penetrates deeper in the inner rim of the disc, ripping off much more gas, and the torque is simply proportional to $\dot{M}$.

Note that, compared to N11 the systems we consider are substantially different:
N11 use 3D SPH with an isothermal equation of state and no selfgravity.
A key difference, is that they find minidiscs around the individual BH, which we do not.
This must be attributed to the fact, that our gas is allowed to cool proportional to the local dynamical time scale,
whereas their simulations are globally isothermal.
Nonetheless the BHB-disc interaction is described by  pure gravitational model, 
which can be applied to a large variety of situations, independently on the particular structure of the disc.

\subsection{Prograde run}
We put forward a 'dust' model also for the prograde case. 
As a matter of fact, Lindblad resonances are purely gravitational effects reflecting
the fact that no stable hierarchical triplets exist in Newtonian dynamics unless the ratio 
of the semi--major axis of the inner and the outer binary is larger than a certain value.
In our case we can think of the BHB as the inner binary and a gas particle in the disc as the outer body orbiting
the center of mass of the BHB.
The transition between stable and unstable configurations corresponds to the location of the inner edge of the cavity.
We can therefore suppose that outside the inner edge, the disc-BHB interaction is weak 
(as certified by the lower left panel in Fig.~\ref{torquecompare}). 
At the inner edge, particles are ripped off and stream toward the binary suffering a slingshot.
In a three body scattering, the average BHB-intruder energy exchange in a single encounter is given by \citep{quinlan96,sesana06}
\begin{equation}
\frac{\Delta{E}}{E}=\frac{2m}{M}C,
\label{deltaEscatt}
\end{equation}
where $E=-\G \mein \mzwe/(2\a)$ is the binary total energy and $m$ is the mass of the intruder.
$C$ is a numerical coefficient of order 1.5, which is determined via extensive scattering experiments.
Recasting equation (\ref{deltaEscatt}) in terms of $\a$ and differentiating, yields
\begin{equation}
\dot{\a}=-2C\frac{\dot{m}}{M}\a.
\label{shrinkscatt}
\end{equation}
However, here $\dot{m}$ is not the accretion rate onto the BHs, but the 'streaming rate' of gas into the cavity, 
flung back to the disc. We give this number in table 2 of paperI\footnote{Note that in table 2 of paperI we give the
streaming rate per orbital period, not per unit time, the conversion factor is simply $2\pi$.},
and  in the adia05 case we get $\dot{m}\approx 5\times10^{-5}$. 
Again, the linear proxy to the solution of equation (\ref{shrinkscatt}) is simply $\a_t=\a_0[1-2C\dot{m}t/M]$.
If we substitute all the relevant quantities and fix $t=2\pi\times45$, we get $\a_{45}\approx0.965\a$. 
This is about $30\%$ larger than the observed shrink, which is acceptable for such a simple model. 
The origin of the discrepancy is likely due to dissipation not taken into account here. 
Part of the streaming gas is actually captured by the hole: in this case, 
for a particle of mass $m$ the binary acquires a binding energy $-\G Mm/(2R_{\rm cavity})$ 
rather than losing an energy given by equation \ref{deltaEscatt} (i.e.,
$\G m\mmu C/\a$). In our case, $\mmu\approx 0.2$. so that the two terms are comparable.
Given that $\approx25\%$ of the streaming material is accreted, this accounts for the discrepancy.
This model also provides an intuitive explanation of the negative torque building up in the region $\a<r<2\a$ in the prograde case.
The energy exchange given by equation (\ref{deltaEscatt}) corresponds to a velocity increment 
of the gas of the order $\Delta{v}_{\gas}\approx C\sqrt{2\G \mmu/(M\a)}$.
Providing a torque on the gas $T_{\gas}=(dL/dt)_{\gas}\approx-\dot{M}\Delta{v}_{\gas}\a$
(we assume an impulsive interaction at $r=\a$ and ${\bf v}_{\gas} \cdot {\bf r}=0$).
This gives, in simulation units a torque acting on the binary of the order
$T_{\rm BHB}=(dL/dt)_{\rm BHB}=-(dL/dt)_{\gas}\approx2\times10^{-5}$,
as observed in the lower left panel of Fig.~\ref{torquecompare}.

In our simulations, therefore, it makes no significant difference to the shrinking rate whether the BHB is aligned or antialigned 
with its disc. However this appears to be a coincidence residing in the ratio of the ripped-off gas in the co--rotating case,
versus the accreted gas in the retrograde case. Our purely dynamical model does not provide an explanation for this,
which stems from the hydrodynamical properties of the gas flow. However, it can be used to quickly determine the evolution 
of a binary once the in-streaming and accretion rates are known.


\subsection{Accretion rates}
We turn now to the observed trends in the accretion rates highlighted in Fig.~\ref{countmm}. Those can be understood by comparing the effect of the mass ratio \q\, and \ecc\, on the mean distance and relative velocity of the two BHs with respect to the gas at the disc edge. We expect the primary to start accreting much more efficiently as soon as its influence radius at apocenter is equal to the apocenter of the secondary, since in this case the secondary can no longer completely shield the primary from the infalling material.  At apocenter, the distances of the BHs to the CoM are $r_i= \a(1+\ecc)/(1+M_i/M_{j})$. If indeed accretion is Bondi--like, then the influence radius is given by: $r_{\rm{a}} \equiv \G M/(v_{\infty}^2+{\rm c}_{s,\infty}^2)$, where $v_{\infty}=v_{\rm i}+v_{\rm edge}$ and the speeds  at apocenter are
\beq
v^2_2=\frac{\G M_1}{(\q+1)\a}\left(\frac{2}{1+\ecc}-1\right) ;\,\, v^2_1=\frac{\G M_2 \q}{(\q+1)\a}\left(\frac{2}{1+\ecc}-1\right).
\eeq
If we impose a turnover in the relative accretion when $r_1(\ecc)+r_{\rm{a}_1}(\ecc)= r_2(\ecc)$, and we solve for $\ecc$, we get:
\beq
\ecc|_{\rm t}=
\frac{ \a c_s^2 (\mzwe^2-\mein^2) + \G (\mein^2 \mzwe  +  \mein \mzwe^2 + 2 \mzwe^3)}{ [\a c_s^2 (\mein +\mzwe)+ \G \mein(\mein + 2  \mzwe)](\mein - \mzwe)}, \\\nonumber
 \eeq
which for the average sound speed of our disc, $c_s\approx0.2$, and $\a\approx1$ gives
\beq
\ecc|_{\rm t}\sim 0.45.
\eeq
We should mention that the dependence on sound speed, $c_s$, is not very strong as long as the discs are supersonic. Our discs  have Mach numbers between 5 and 10. Additionally, the fact that in the retrograde case for $\ecc \gsim 0.8$ the ratio $\dot{M_2}/\dot{M_1}$ becomes smaller than unity can be attributed to the fact that the binary leaves the disc plane and starts tilting, so the gas can come closer to the primary without being captured by the secondary.

\section{Alignment between the BHB and the disc plane: the tilting instability}
\label{sec:tilting}
When looking at the bottom right panel Fig.~\ref{countlow}, right, we find that the inclination of the BHB tilts mildly away from exact antialignment ($\i=180$) in the ${\tt retro\, \ecc=0.8}$ run. At the same time  \ecc\,  saturates its growth and turns over to become again more circular. In order to study this effect in more detail, we setup the additional run $"{\tt retro^{**} \ecc=0.9}"$ with two times the resolution of the default runs. Additionally, we monitored the numerical drift of the total angular momentum vector ${\bf L}$ to be small and, averaged over an orbit, consistent with a numerical zero. Also, all conservation laws are monitored very closely, and we stop trusting the evolution of $"{\tt retro^{**} \ecc=0.9}"$ after orbit $600$. 

As indicators of the onset of tilting, we identify the following criteria observed in the low-resolution runs:
\begin{itemize}
 \item the ratio of the individual accretion rates would drop below $\ratio<1$; 
 \item the {\it outward} torque would become significantly positive;
 \item the individual differential $z-$torques onto the BHs would stop alternating sign between each peak
 and rather be pairwise {\it in phase};
 \item the angular momentum (modulus) of the relevant parts of the disc be larger than that of the BHB, i.e. $L_{\rm Disc}>L_{\rm BHB}$.
\end{itemize}
We define  the relevant angular momentum of the disc to be 
\beq
L_{\rm Disc} \approx \sum_i ({\bf r}_i-{\bf r}_{\rm CoM}) \times {\bf p}_i  \,\, , |{\bf r}_i-{\bf r}_{\rm CoM}| < 2.5 \a(1+\ecc).
\label{ldisc}
\eeq
Because outwards of this radius, the amplitude of the individual differential torque is essentially zero{\footnote{Looking at the $t=1332$ row in Fig. \ref{fig:tilting}, we note that this radius is indeed compatible with the 'warping radius' at which the disc breaks in several differentially precessing annuli. The portion of the disc inside this raduis can actively respond to the BHB torques, and is therefore the relevant one for the computation of $L_{\rm Disc}$ \citep{volo07}}}.
While it is true that in order to transport outwards the angular momentum, a radially larger disc is required, 
the dynamical influence onto the BHB, i.e. the strength and orientation of the torque is given only by the mass and distance
of the matter that orbits at radii where the torque is non-zero.
For $e>0.9$ the angular momentum of the BHB drops below $0.5L_{\rm Disc}$, which is the condition under which a
counteraligned stable configuration is not possible anymore \cite{King2005}. 
At this point, it should be mentioned, that the two CoM, the one of the BHB and the global CoM exhibit 
increasing amount of relative movement with \ecc. 

\begin{figure}
\centering
\includegraphics[scale=0.4,clip=true,angle=0]{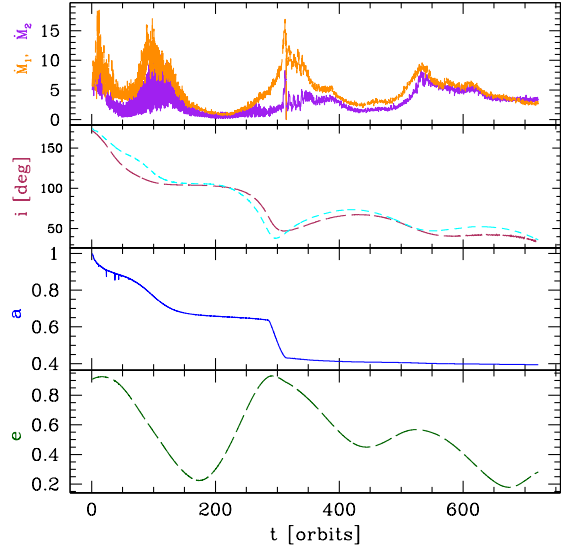}
\caption[Retrograde high eccentricity run, different accuracies]{
{\sc Retrograde high eccentricity run, tilting of the BHB } - The $"{\tt retro^{**} \ecc=0.9}"$ run; from the top to the bottom, panels show the individual accretion rates \textcolor{orange}{$\dot{M}_1$}, \textcolor{RoyalPurple}{$\dot{M}_2$} ($\times 1000$), the \textcolor{Maroon}{total inclination $\i$} in degrees, the \textcolor{cyan}{inclination ${\rm i_{2.5}}$ towards only the inner rings}, \textcolor{blue}{semi-major axis $\a$} and the \textcolor{OliveGreen}{eccentricity $\ecc$}. See text for more details; quantities are in code units.
} 
\label{tilting}
\end{figure}

\subsection{Physical reasons for the tilting}
The retrograde setup differs in several physical properties from the prograde, such that the BHB grazes the disc without the ``stream bending'' within the cavity. The appearance of minidiscs is suppressed in the retrograde case, because firstly the cavity is too small to accommodate the discs and secondly because the gas inflows onto the BHs is much more radial than in the prograde case. So the gas is accreted mostly in the wake of the BHs in the retrograde case. Every direct BHB-disc interaction is thus much stronger than in the prograde case, but takes place in more closely confined space. As \ecc\, increases (and with it $L_{\rm Disc}/L_{\rm BHB}$) this induces an increasingly large relative movement between the CoM of the BHB  with respect to the global CoM. It should be mentioned that the retrograde disc remains circular even if the BHB has high $\ecc\sim 0.8$ (cf. Fig.~\ref{fig:tilting}, top panel). Thus, it can be understood that the BHB, which has increasingly small angular momentum, has less and 
less rotational support feeling the impacts into its disc more strongly with increasing $\ecc$. The details for a BHB to become secularly unstable to this tilting instability will strongly be  disc--model dependent, which is why we only list under which conditions we expect a system to go unstable:
\begin{enumerate}
 \item a retrograde disc of $\i=180$\textdegree\, is an unstable equilibrium configuration if $L_{\rm Disc}> 2L_{\rm BHB}$ as is the case here: with growing $\ecc$ the BHB is increasingly interacting with the retrograde disc, it is not off-hand to expect perturbations that are not perfectly  in--plane;
 \item the corotating case is stable attractor, there should be a threshold value of perturbation that leads the BHB to leave the retro-configuration in favor of the prograde one;
 \item if the accretion disc is not perfectly cylindrically symmetric, but rather clumpy and of non-zero vertical extent, perturbations out-of-plane \textit{and} out-of-synchronization\footnote{Out-of-synchronization means that the accretion events do not happen coincidentally at the two BHs} are well possible.
\end{enumerate}
It is thus not surprising that simulations where discs are light and self-gravity
is not considered \citep[e.g,][]{nixon2012,nixon13} would not find such an instability.

\subsection{The stages of tilting}
We now further explore the tilting itself, assuming that it is a physical effect.
The time evolution of this run covers almost $3000$ orbits (when taking into account the very significant shrinkage of \a(t)).
After an initial increase of $\ecc$ there follows circularisation which again is followed by an increase in $\ecc$.
At the same time the inclination \i\, is decreasing then saturating and oscillating until it finally decreases. \a\,
experiences several phases of strong shrinkage\footnote{Note, that while energy of the binary shrinks 
drastically, this is not due to numerical integration error but to the cooling function imposed on the gas. We have verified
that at any given moment in the simulation, the numerical energy errors are (about a factor $10$)
smaller than the prescribed energy loss of the system.}, whereas the accretion rate onto the primary exceeds that onto the secondary
until very late times.
This behavior of exchanging inclination and 
eccentricity is somewhat reminiscent of the Kozai--effect\footnote{The Kozai effect \citep{Kozai} is a well known,
well studied phenomenon of a secular three-body interaction when a binary interacts with a perturber that is on an
inclined orbit with respect to the binary plane. The perturber can then oscillate in its inclination, pericentre,
longitude of ascending node and eccentricity \citep{Genya2008} in so-called Kozai-cycles. 
The perturber feels secularly torqued by the binary and starts precessing. 
Depending on the time-scale of this precession, the binary eccentricity can be excited. 
For four-body systems, i.e. a binary and two perturbers, the same effect can be found \citep[e.g.][]{Murray1999} 
where now the two perturbers can additionally exchange inclination and eccentricity.}. 
However, it is not straightforward to apply these few body calculations for the BHB-disc 
scenario studied here: a gaseous disc is not a rigid body and can dissipate angular momentum; nor is the disc a well located point mass 
with clearly defined eccentricity and inclination in itself.

\begin{figure*}
\centering
\includegraphics[width=1\linewidth]{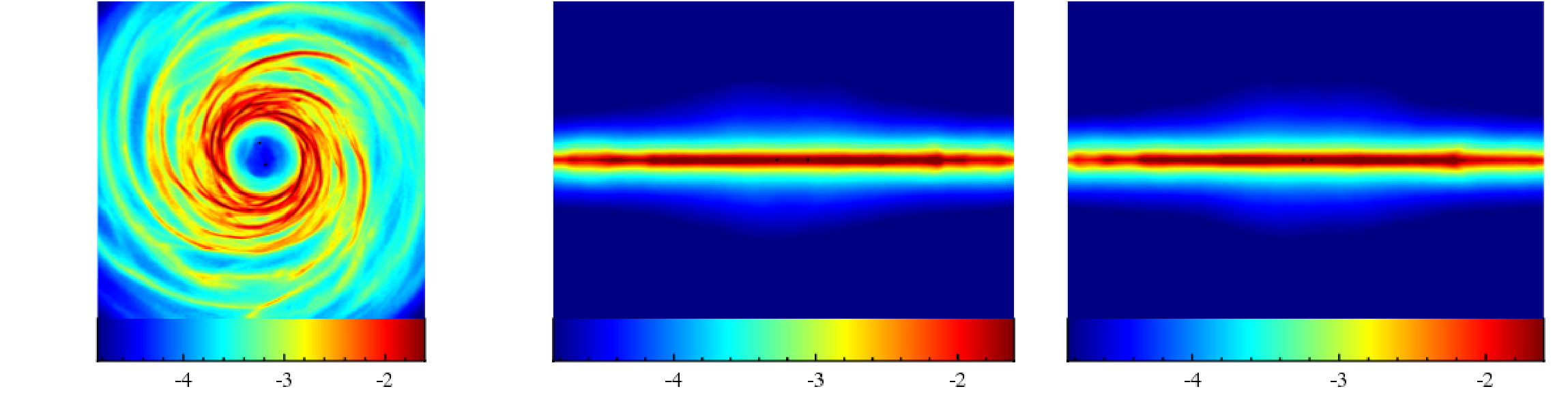}
\includegraphics[width=1\linewidth]{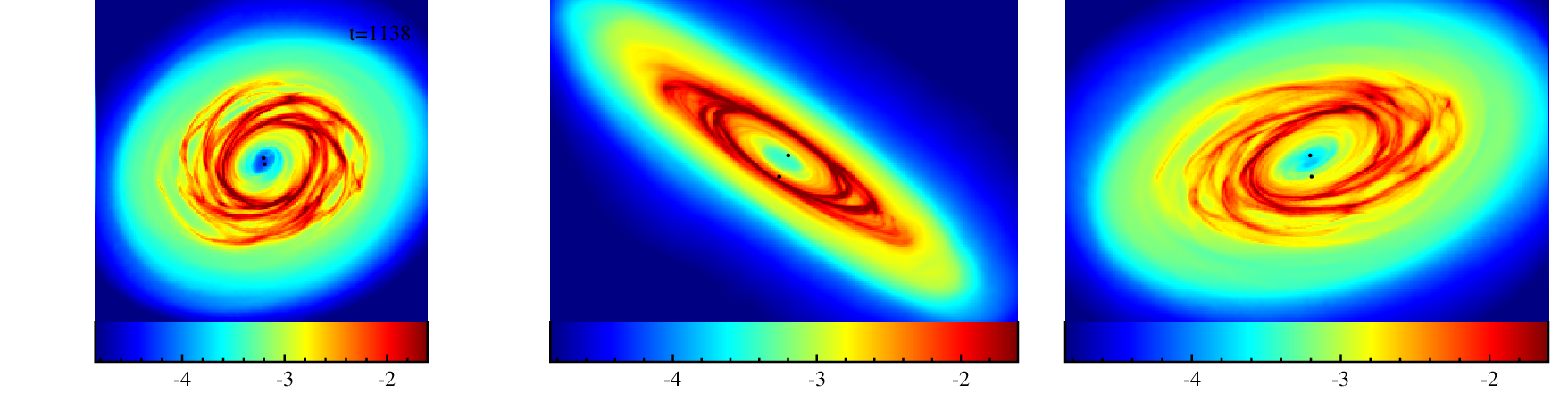}
\includegraphics[width=1\linewidth]{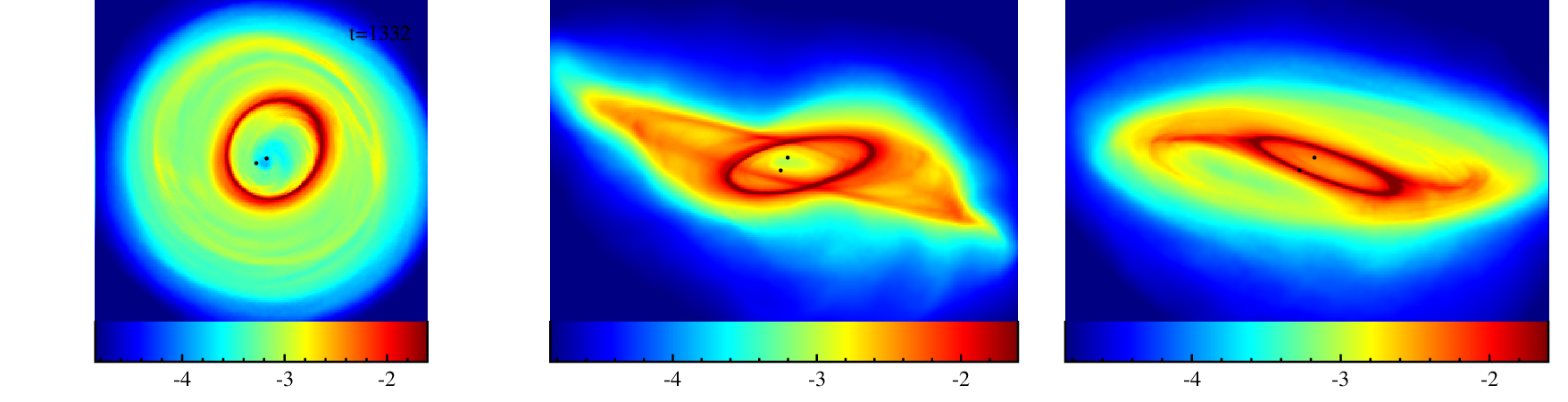}
\includegraphics[width=1\linewidth]{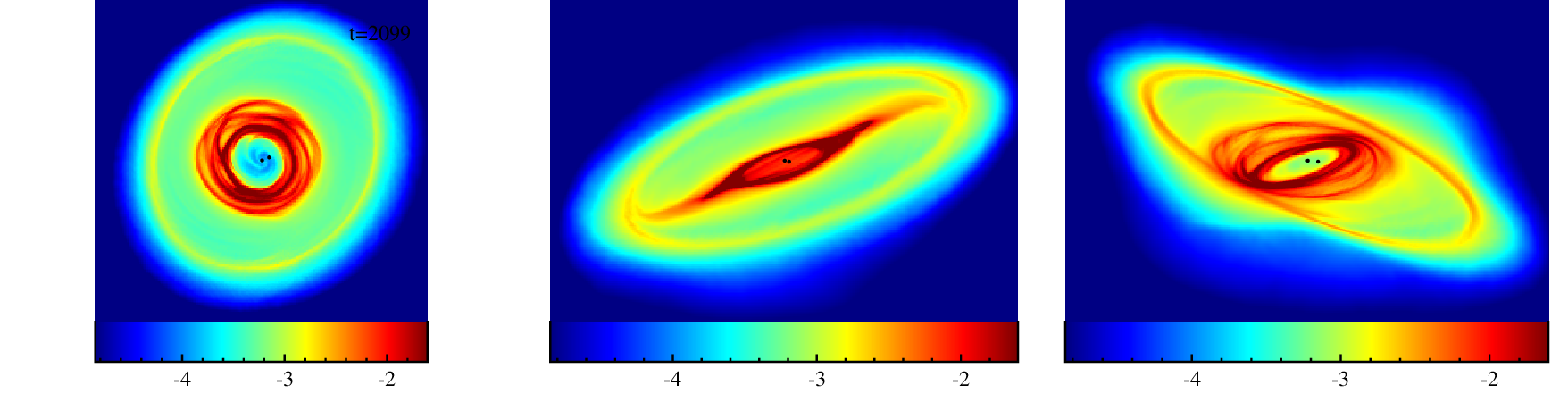}
\includegraphics[width=1\linewidth]{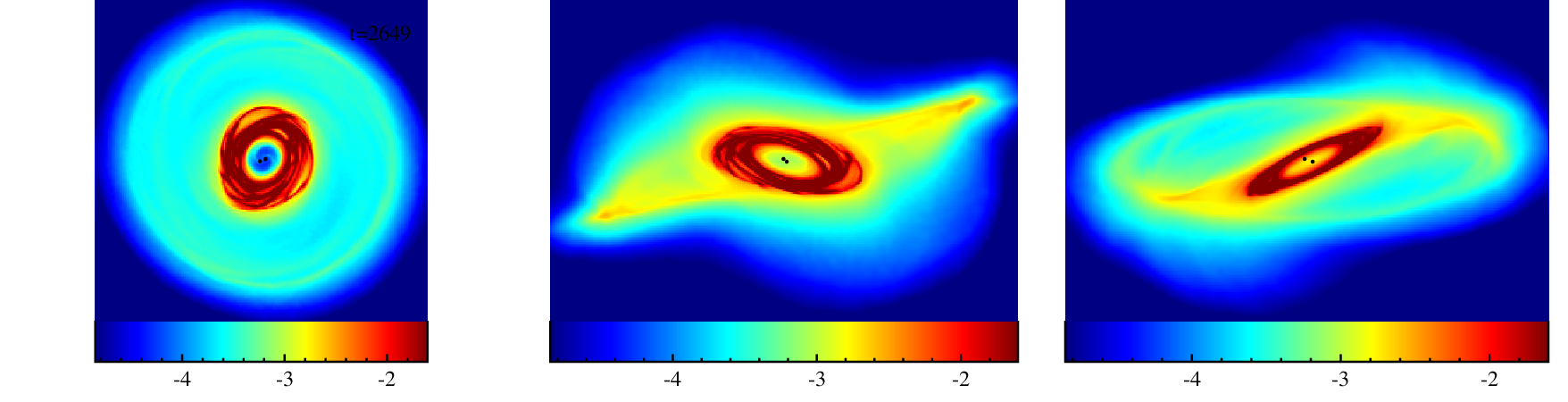}
\caption{
{\sc The tilting instability: logarithmic column density, face--on $xy$, edge--on $xz$, edge--on $yz$ } - Time evolution (top to bottom) of the high accuracy, tilting $"{\tt retro^{**}\ecc=0.9}"$ run. Panels show the successive stages of the binary first tilting out of plane and then subsequently aligning with the disc. Note the spread into various annuli that the disc undergoes in the middle panels.
}
\label{fig:tilting}
\end{figure*}

The stages that the BHB undergoes on its way towards alignment can be subdivided into the following which are supported
by Fig.~\ref{tilting} and \ref{fig:tilting}:
\begin{enumerate}
 \item \textit{Fast circularisation}: the BHB tilts out of plane, $\ecc$ quickly shrinks to $\ecc\sim 0.2$, while
at the same time \a\, shrinks almost by a factor of two.
 \item \textit{Disc fallback I}: while the BHB has shrunk very fast \textit{and} circularized \textit{and} tilted out of plane, the disc  had disintegrated into many annuli of various inclinations due to conservation of angular momentum. Now, however, the inner part of the gas forms a flat inner disc again. Meanwhile, the gravity torques from the precessing gaseous annuli cause the BHB eccentricity to grow. It should be noted, that the CoM of the BHB and the (inner) disc have drifted apart significantly by now.  The accretion rates are low and the semi-major axis remains almost stationary.
 \item \textit{Inner asymmetric ring collapse}: once the gas has reformed into one (inner) planar disc, the new disc-configuration has lost the formerly large BHB upkeeping the cavity size. So, at this point, it re-adjusts to the new BHB ``size'' and a hot, dense ring forms (cf. third panel Fig.~\ref{fig:tilting}). This ring has very low Toomre parameter, becomes extremely clumpy and the CoM of BHB and disc-ring drift even further apart. Note, that the cooling in the ring depends on the local orbital timescale. The fallback of gas is accompanied by large accretion rates and thus induced semi-major axis shrinkage. The BHB eccentricity remains very high.
 \item \textit{Circularisation II}: when finally the ring was able to cool enough for its inner part to reach the BHB, the BHB starts circularizing and tilting \textit{into} the disc. The resulting meta-stable configuration has an inclination of about $\i\sim 90$\textdegree\,, is mildly eccentric, a stable semi-major axis and comparatively low accretion rates.
 \item \textit{Disc fallback II}: the initial disc has left behind some outer annuli that now start interacting with the outer part of the (inner) disc. The inner disc and the outer ring are now aligning which is accompanied by a series of precessions of the outer ring. Meanwhile the BHB increases its $\ecc$ and stays in the same orientation while the disc tilts towards the BHB.
 \item \textit{Outer ring collapse}: Similar to stage (3), now the outer ring becomes unstable under its own gravity as it has lost too much rotational energy by aligning with the inner disc. (Most of) the outer ring collapses onto the inner disc. The accretion rates are high, but $\a$ only mildly shrinks.
 \item \textit{CoM alignment}: Once the inner disc and the BHB are the only two major pieces reformed and the inclination is below $60$\textdegree, the BHB and its new, much more compact disc start aligning their CoMs. This is accompanied by a circularisation of the BHB, as during  the interaction  the BHB pushes the disc towards one side by impacting into it.
 \item \textit{Standard $\ecc$-growth}: when finally the BHB and the disc have re-united their CoMs (apart from a tenuous outer ring), the excitation of $\ecc$ similar to what was described in \citet{Roedig2011} starts occurring, while the inclination furthermore decreases.
\end{enumerate}

At the final state of the simulation, the BHB has an inclination of $\i\sim 30$\textdegree\,, an eccentricity of about $\ecc\sim 0.3$ and the secondary accretion rate is exceeding the primary. During this evolution, the drift of the total angular momentum vector was about $0.8$\textdegree, and the  $\beta-$cooling has induced a loss of $L$ of about $1.2\%$. However, we do not trust the evolution past orbit $600$, because there appears noise in both the total energy and the total angular momentum and the ratio between $r_{\rm sink}$ and \a\, are severely weakening the ${\it inwards}$ torque, thus vitiating the secular effects drastically. However, from a physical point of view, the aligned, prograde state should be energetically preferred and  once the disc has aligned with the BHB, it will be forced to expand its cavity analogously to the prograde case. The numerical problem of performing several thousands of orbits is challenging and we argue here that studying the entire final alignment up to $\ecc\sim 0.
6$ (cf. \cite{Roedig2011}) would be a waste of CPU time, since the physical arguments are quite strong 
if indeed the instability occurs in the first place. Also, 
it is our suspicion that different choices for $\beta$, cooling prescriptions, 
and global disc thermodynamics will alter the 8 stages identified here. 
The cooling, i.e. the loss of angular momentum seems to be the most crucial numerical parameter
in setting the timescales for the ``ring collapse'' and the dissipation of the several annuli.

\section{Summary}
\label{discussion}

We presented numerical work on the evolution of the orbital elements of a comparable
mass black hole binary surrounded by a retrograde selfgravitating disc.
First, it was found that for quasi-circular BHB-disc configurations,
the difference between the retrograde and prograde  scenario is chiefly in the eccentricity and mass ratio  evolution. 
The eccentricity remains close to zero in the retrograde case whereas it grows in the prograde case. 
The semi-major axis shrinkage was found to be the same for the circular BHB irrespective of disc (anti)alignment. 
For increasingly eccentric set-ups, the growth of $\ecc$ was found to be exponential in time and the coupled evolution 
of the semi--major axis $\dot{\a}$ was observed to be proportional to \ecc(t), thus making the decay very efficient.
Additionally, it was found that the individual accretion rates onto the two BHs grow, but that their ratio secondary over
primary $\ratio$ assumes a maximum at eccentricities around $\ecc\sim 0.45$. 

We demonstrated that all the observed features can be satisfactorily explained using a simple 'dust' model in which the BHB-disc
interaction is purely gravitational. In the retrograde case, the absence of resonances together with the high BH-disc 
relative velocity (i.e., a small cross section) imply that the gas interacts with the binary only via direct impact onto the secondary.
This is confirmed by the fact, that the turnover of $\ratio(\ecc)$ could be explained by Bondi-Hoyle like accretion assuming ballistic speeds
of disc-edge and binary at apocenter.
Generalizing the model of N11, we find that this simple 'impact model' is able to catch of the relevant features of the binary 
evolution at different eccentricities. In the prograde case, the evolution is determined by the gas inflows flung back to the disc.
Treating the gas as a dust particle, we found that a simple model based on three body scattering theory fully accounts
for the BHB-disc interplay. In those models, however, the mass accretion rate and instreaming rate stem from the hydrodynamical
properties of the gas flow and are input parameters.

The retrograde setup shows non-negligible  relative motion between the BHB center of mass and the global center of mass (CoM),
even in circular runs. In the prograde case, this had been a very small effect only.
There are indications that the BHB eccentricity will not grow to $\ecc\sim 1$ in the retrograde case,
but that the impacts into the disc inner edge destabilize the BHB until it starts moving out of the binary-disc plane.
The most important physical conditions for this instability to occur be that the angular momentum of the disc exceed twice
that of the BHB and that the disc be sufficiently non--axisymmetric for out-of-plane perturbations to occur. 
The details of the subsequent alignment are likely to be sensitive to the cooling prescription of the disc,
but it is to be expected that the disc disintegrates into several annuli of various inclinations which precess within each other.
It was observed that precession of the disintegrated disc annuli causes the BHB eccentricity to grow while the BHB CoM 
and the disc CoM do not overlap. There were stages wherein the multiple annuli aligned again to try form a coherent, flat disc.
Due to loss of rotational support from the shrunk BHB, this was preceded  by a short period of fragmentation-like behavior where the disc 
collapsed shortly into a dense ring, to spread out into spiral arms later. The BHB eccentricity exhibited few oscillations and it is expected 
that the final state will be a perfectly aligned disc-BHB system of eccentricity around $\ecc\sim 0.6$.

We note, that it is well conceivable that, depending on the physical separation and mass, during any point of the highly eccentric phases
the BHB might enter the relativistic regime and rapidly lose energy via gravitational radiation. At this point the circularization effect of gravitational radiation would compete with the eccentricity growth enforced by the retrograde disc, possibly causing a distinctive eccentricity evolution of the system that would be possibly detectable and discernible by low--frequency gravitational wave interferometers \citep{amaro13a,whitepaper13}, offering an interesting observational test of the BHB-counterrotating disc coupling.

\section{Acknowledgments}
CR is indebted to Kareem Sorathia and Sarah Buchanan without whom this paper would not have been written.
CR is grateful  to Julian Krolik and Massimo Dotti for helpful suggestions and was 
happy to have  informative conversations with  Cole Miller, Sanchayeeta Borthakur and Jens Chluba. 
All computations were performed on the {\it datura} cluster of the AEI and CR thanks Nico Budewitz
for his HPC support. This work was funded in part by the 
International Max--Planck Research School, NSF grants AST-0908336, AST-1028111 and the SFB Transregio 7.

\bibliography{aeireferences.bib,local.bib}
\end{document}